\documentclass[aps,floatfix,prd,nofootinbib,superscriptaddress,reprint,showpacs,10pt,preprintnumbers,longbibliography]{revtex4-1}
\usepackage[utf8]{inputenc}
\usepackage[pdftex]{graphicx}
\usepackage{float}
\usepackage{amsmath}
\usepackage{amssymb}
\usepackage{mathtools}
\usepackage{siunitx}
\usepackage{amsfonts}
\usepackage{dsfont}
\usepackage{array}
\usepackage{bm}
\usepackage{mathrsfs}
\usepackage{pifont}
\usepackage{multirow}
\usepackage{upgreek}
\usepackage[dvipsnames]{xcolor}
\usepackage[pdftex,
  pdftitle={On the generalised eigenvalue method and its relation to
    Prony and generalised pencil of function methods}
  pdfauthor={Matthias Fischer, Bartosz Kostrzewa, Johann
    Ostmeyer, Konstantin Ottnad, Martin Ueding, Carsten Urbach},
  bookmarks,
  colorlinks,
  linkcolor=myblue,
  citecolor=mymagenta,
  menucolor=black,
  urlcolor=myblue,
  plainpages=false,
  pdfpagelabels,
  hypertexnames=false]{hyperref}
\usepackage{verbatim}
\usepackage{slashed}
\usepackage{cleveref}
\usepackage{ucs}
\usepackage{subfigure}

\newcommand{\fig}[1]{Figure~\ref{#1}}
\newcommand{\tab}[1]{Table~\ref{#1}}
\newcommand{\eq}[1]{Eq.~(\ref{#1})}

\renewcommand{\l}{\left}
\renewcommand{\r}{\right}

\DeclareMathOperator{\im}{i}
\DeclareMathOperator{\ord}{\mathcal{O}}
\newcommand{\eto}[1]{\ensuremath{\mathrm{e}^{#1}}}
\newcommand{\md}{\ensuremath{\mathrm{d}}}
\newcommand{\ordnung}[1]{\ensuremath{\ord\left(#1\right)}}
\newcommand{\gm}{Gardner method}
\newcommand{\Meff}{M_\mathrm{eff}}
\newcommand{\tMeff}{\tilde{M}_\mathrm{eff}}

\definecolor{mymagenta}{RGB}{200, 0, 100}
\definecolor{myblue}{RGB}{45, 48, 146}

\graphicspath{{plots/}}

\begin{document}
\title{On the generalised eigenvalue method\\ and its relation to Prony
and generalised pencil of function methods}
\author{Matthias Fischer}
\affiliation{Helmholtz-Institut f\"ur Strahlen- und Kernphysik, University of Bonn, Nussallee 14-16, 53115 Bonn, Germany}
\affiliation{Bethe Center for Theoretical Physics, University of Bonn, Nussallee 12, 53115 Bonn, Germany}
\author{Bartosz Kostrzewa}
\affiliation{Helmholtz-Institut f\"ur Strahlen- und Kernphysik, University of Bonn, Nussallee 14-16, 53115 Bonn, Germany}
\affiliation{Bethe Center for Theoretical Physics, University of Bonn, Nussallee 12, 53115 Bonn, Germany}
\author{Johann Ostmeyer}
\affiliation{Helmholtz-Institut f\"ur Strahlen- und Kernphysik, University of Bonn, Nussallee 14-16, 53115 Bonn, Germany}
\affiliation{Bethe Center for Theoretical Physics, University of Bonn, Nussallee 12, 53115 Bonn, Germany}
\author{Konstantin Ottnad}
\affiliation{PRISMA$^+$ Cluster of Excellence and Institut f\"ur Kernphysik, Johann-Joachim-Becher-Weg~45, University of Mainz, 55099 Mainz, Germany}
\author{Martin Ueding}
\affiliation{Helmholtz-Institut f\"ur Strahlen- und Kernphysik, University of Bonn, Nussallee 14-16, 53115 Bonn, Germany}
\affiliation{Bethe Center for Theoretical Physics, University of Bonn, Nussallee 12, 53115 Bonn, Germany}
\author{Carsten Urbach}
\affiliation{Helmholtz-Institut f\"ur Strahlen- und Kernphysik, University of Bonn, Nussallee 14-16, 53115 Bonn, Germany}
\affiliation{Bethe Center for Theoretical Physics, University of Bonn, Nussallee 12, 53115 Bonn, Germany}
\collaboration{Extended Twisted Mass Collaboration}
\date{\today}
\begin{abstract}
  We discuss the relation of three methods to determine energy levels
  in lattice QCD simulations: the generalised eigenvalue, the Prony
  and the generalised pencil of function methods. All three can be
  understood as special cases of a generalised eigenvalue problem.
  We show analytically that the leading corrections to an energy $E_l$
  in all three methods due to unresolved states decay asymptotically
  exponentially like $\exp(-(E_{n}-E_l)t)$. Using synthetic data we
  show that these corrections behave as expected also in practice. We
  propose a novel combination of the generalised eigenvalue and the
  Prony method, denoted as GEVM/PGEVM, which helps to increase the energy
  gap $E_{n}-E_l$. We illustrate its usage and performance using
  lattice QCD examples.  
\end{abstract}

\maketitle

\section{Introduction}

In lattice field theories one is often confronted with the task to
extract energy levels from noisy Monte Carlo data for Euclidean
correlation functions, which have the theoretical form
\begin{equation}
  \label{eq:corr}
  C(t)\ =\ \sum_{k=0}^\infty\ c_k\, e^{-E_k t}
\end{equation}
with real and distinct energy levels $E_{k+1}>E_k$ and real coefficients
$c_k$. It is well known that this task represents an ill-posed problem
because the exponential functions do not form an orthogonal system of
functions.

Still, as long as one is only interested in the ground state $E_0$
and the statistical accuracy is high enough to be able to work at
large enough values of $t$, the task can be accomplished by making use
of the fact that
\begin{equation}
  \label{eq:limit}
  \lim_{t\to\infty} C(t)\ \approx\ c_0 e^{-E_0 t}\,,
\end{equation}
with corrections exponentially suppressed with increasing $t$ due to ground 
state dominance. However, in lattice quantum chromodynamics, the
non-perturbative approach to quantum chromodynamics (QCD), the signal 
to noise ratio for $C(t)$ deteriorates exponentially with increasing
$t$~\cite{Lepage:1989}. Moreover, at large Euclidean times there 
can be so-called thermal pollutions (see e.g.\ Ref.~\cite{Feng:2010es})
to the correlation functions, which, if not accounted for, render the
data at large $t$ useless. And, once one is interested in excited
energy levels  $E_k\,,\ k>0$, alternatives to the ground state
dominance principle need to be found.

The latter problem can be tackled applying the so-called generalised
eigenvalue method (GEVM) -- originally proposed in
Ref.~\cite{Michael:1982gb} and further developed in
Ref.~\cite{Luscher:1990ck}. It is by now well established in lattice
QCD applications and allows one to estimate ground and excited states
for the price that a correlator matrix needs to be computed instead of
a single correlation function. Moreover, the systematics of this
method are well understood~\cite{Luscher:1990ck,Blossier:2009kd}. 

An alternative method, originally proposed by de
Prony~\cite{Prony:1795}, represents an algebraic method to determine
in principle all the energy levels from a single correlation
function. However, it is well known that the Prony method can become
unstable in the presence of noise. The Prony method was first used for
lattice QCD in Refs.~\cite{Fleming:2004hs,Beane:2009kya}. For more
recent references see
Refs.~\cite{Fleming:2006zz,Berkowitz:2017smo,Cushman:2019hfh} and also
Appendix~\ref{sec:gardner}. For an application of the Prony method
in real time dynamics with Tensor networks see Ref.~\cite{Banuls:2019qrq}.

In this paper we discuss the relation among generalised eigenvalue,
Prony and generalised pencil of function (GPOF) methods and trace them all
back to a generalised eigenvalue problem. This allows us to derive the
systematic effects due to so-called excited state contributions for
the Prony and GPOF methods using perturbation theory invented for the
GEVM~\cite{Blossier:2009kd}. In addition, we propose a combination of
the GEVM and the Prony method, the latter of which we also formulate
as a generalised eigenvalue method and denote it as Prony GEVM (PGEVM).
The combination we propose is to apply first the GEVM to a correlator
matrix and extract the so-called principal correlators, which are
again of the form \eq{eq:corr}. Then we apply the PGEVM to the principal
correlators and extract the energy levels. In essence: the GEVM is
used to separate the contributing exponentials in distinct principal
correlators  with reduced pollutions compared to the original
correlators. Then the PGEVM is applied only to obtain the ground state
in each principal correlator, the case where it works best.

By means of synthetic data we verify that the PGEVM works as expected
and that the systematic corrections are of the expected form. 
Moreover, we demonstrate that with the combination GEVM/PGEVM example
data from lattice QCD simulations can be analysed: we study the pion
first, where we are in the situation that the ground state can be
determined with other methods with high confidence. Thereafter we also
look at the $\eta$-meson and $I=1, \pi=\pi$ scattering, both of which
require the usage of the GEVM in the first place, but where also noise
is significant. 

The paper is organised as follows: in the next section we introduce the GEVM 
and PGEVM and discuss the systematic errors of PGEVM. After briefly explaining 
possible numerical implementations, we present example applications using both 
synthetic data and data obtained from lattice QCD simulations. In the end we 
discuss the advantages and disadvantages of our new method, also giving an 
insight into when it is most useful.

\section{Methods}
\label{sec:method}

Maybe the most straightforward approach to analysing the correlation
function \eq{eq:corr} for the ground state energy $E_0$ is to use the
so-called effective mass defined as
\begin{equation}
  \label{eq:effmass}
  M_\mathrm{eff}(t_0, \delta t)\ =\ -\frac{1}{\delta t}\log\left(
  \frac{C(t_0+\delta t)}{C(t_0)}\right)\,.
\end{equation}
In the limit of large $t_0$ and fixed $\delta t$, $M_\mathrm{eff}$
converges to $E_0$. The correction due to the first excited state
$E_1$ is readily computed:
\begin{equation}
  \begin{split}
    M_\mathrm{eff}(t_0, \delta t)\ \approx&\ E_0 + \frac{c_1}{c_0}
    e^{-(E_1-E_0)t_0}\\
    &\times \left( 1 - e^{-(E_1-E_0)\delta t}\right)\,\frac{1}{\delta t}\,.
  \end{split}
\end{equation}
It is exponentially suppressed in $t_0$ and the energy difference between
first excited and ground state. It is also clear from this formula
that taking the limit $\delta t\to\infty$ while keeping $t_0$ fixed
leads to a worse convergence behaviour than keeping $\delta t$ fixed
and changing $t_0$. In this section we will 
discuss how both of the two above equations generalise.

\subsection{The generalised eigenvalue method (GEVM)}

We first introduce the GEVM. The method is important for being able to
determine ground and excited energy levels in a given
channel. Moreover, it helps to reduce excited state contaminations to
low lying energy levels. 

Using the notation of Ref.~\cite{Blossier:2009kd}, one considers
correlator matrices of the form
\begin{equation}
  \label{eq:corrM}
  C_{ij}(t)\ =\ \langle \hat O_i^{}(t')\ \hat O_j^\dagger(t'+t)\rangle\ =\ \sum_{k=0}^\infty e^{-E_k t} \psi_{ki}^*\psi_{kj}\,,
\end{equation}
with energy levels $E_k > 0$ and $E_{k+1} > E_k$ for all values of
$k$. The $\psi_{ki}=\langle 0|\hat O_i|k\rangle$ are matrix elements of
$n$ suitably chosen operators $\hat O_i$ with $i=0, ..., n-1$. Then, the
eigenvalues or so-called principal 
correlators $\lambda(t, t_0)$ of the generalised eigenvalue problem (GEVP)
\begin{equation}
  C(t)\, v_k(t, t_0)\ =\ \lambda^0_k(t, t_0)\, C(t_0)\, v_k(t, t_0)\,,
  \label{eq:GEVP}
\end{equation}
can be shown to read
\begin{equation}
  \label{eq:lambdaGEVP}
  \lambda^0_k(t, t_0)\ =\ e^{-E_k(t - t_0)}
\end{equation}
for $t_0$ fixed and $t\to\infty$.
Clearly, the correlator matrix $C(t)$ will for every practical
application always be square but finite with dimension $n$. This will
induce corrections to \eq{eq:lambdaGEVP}. The corresponding 
corrections were derived in Ref.~\cite{Luscher:1990ck,Blossier:2009kd}
and read to leading order 
\begin{equation}
  \label{eq:t0corr}
  \lambda_k(t, t_0) = b_k \lambda^0_k(1+\mathcal{O}(e^{-\Delta E_k t}))
\end{equation}
with $b_k>0$ and
\begin{equation}
  \Delta E_k\ =\ \min_{l\neq k}|E_l -E_k|\,.
\end{equation}
Most notably, the principal correlators $\lambda_k(t_0, t)$ are at
fixed $t_0$ again a sum of exponentials. As was shown in
Ref.~\cite{Blossier:2009kd}, for $t_0>t/2$ the leading corrections are
different compared to \eq{eq:t0corr}, namely of order
\begin{equation}
  \label{eq:t0corr2}
  \exp[-(E_n - E_k)t]\,.
\end{equation}

\subsection{The Prony method}

For the original Prony method~\cite{Prony:1795}, we restrict ourselves
first to a finite 
number $n$ of exponentials in an Euclidean correlation function $C^0$
\begin{equation}
  \label{eq:C0}
  C^0(t)\ =\ \sum_{k=0}^{n-1} c_k\, e^{-E_k t}\,.
\end{equation}
The $c_k$ are real, but not necessarily positive constants and $t$ is
integer--valued.  Thus, we focus on one matrix element of the correlator
matrix \eq{eq:corrM} from above or other correlators with the
appropriate form. We assume now $E_k\neq 0$ for all $k\in\{0,
\ldots, n-1\}$ and that all  the $E_k$ are distinct. Moreover, we
assume the order $E_{k+1} > E_k$ for all $k$.
Then, Prony's method is a generalisation of the effective mass
\eq{eq:effmass} in the form of a matrix equation
\begin{equation}
  \label{eq:prony}
  H\cdot x = 0\,,
\end{equation}
with an $(n+1)\times(n+1)$ Hankel matrix $H$ 
\[
  H=
  \begin{pmatrix}
    C^0(t) & C^0(t+1) & \ldots & C^0(t+n) \\
    C^0(t+1) & C^0(t+2) & \ldots & C^0(t+n+1) \\
    \vdots & \vdots & \ddots & \vdots \\
    C^0(t+n) & C^0(t+n+1) & \ldots & C^0(t+2n) \\
  \end{pmatrix}
\]
and a coefficient vector $x = (x_0, \ldots, x_{n-1}, 1)$ of length
$n+1$. After solving for $x$, the exponentials are obtained from $x$
by the roots of  
\[
x_0 + x_1\left(e^{-E_l}\right) + x_2 \left(e^{-E_l}\right)^2 + \ldots
+ \left(e^{-E_l}\right)^n = 0\,. 
\]
For a further generalisation see Ref.~\cite{Beane:2009kya} and
references therein.

\subsection{The Prony GEVM (PGEVM)}

Next we formulate
Prony's method \eq{eq:prony} as a generalised eigenvalue
problem (see also Ref.~\cite{sauer:2013}). Let $H^0(t)$ be a $n\times
n$ Hankel matrix for $i,j=0, 1, 2, \ldots, n-1$ defined by 
\begin{equation}
  \begin{split}
    H^0_{ij}(t)\ &=\ C^0(t + i\Delta + j\Delta)\\
    &=\ \sum_{k=0}^{n-1} e^{-E_kt}\, e^{-E_ki\Delta}\, e^{-E_kj\Delta}
    c_k\,,\\ 
  \end{split}
\end{equation}
with integer $\Delta>0$.
$H^0(t)$ is symmetric, but not necessarily positive definite.
We are going to show that the energies $E_0, \ldots, E_{n-1}$ can be
determined from the generalised eigenvalue problem
\begin{equation}
  \label{eq:hankelgevp}
  H^0(t)\, v_l\ =\ \Lambda^0_l(t, \tau_0)\, H^0(\tau_0)\, v_l\,.
\end{equation}
The following is completely analogous to the corresponding proof of
the GEVM in Ref.~\cite{Blossier:2009kd}. 
Define a square matrix
\begin{equation}
  \chi_{ki}\ =\ e^{-E_k i\Delta}\,.
\end{equation}
and re-write $H^0(t)$ as
\[
H^0_{ij}(t)\ =\ \sum_{k=0}^{n-1} c_k e^{-E_k t}\chi_{ki}\chi_{kj}\,.
\]
Note that $\chi$ is a square Vandermonde matrix
\[
\chi\ =\
\begin{pmatrix}
  1 & e^{-E_0\Delta} & e^{-2E_0\Delta} & \ldots & e^{-(n-1)E_{0}\Delta}\\
  1 & e^{-E_1\Delta} & e^{-2E_1\Delta} & \ldots & e^{-(n-1)E_{1}\Delta}\\
  \vdots & \vdots & \vdots & \ldots & \vdots \\
  1 & e^{-E_{n-1}\Delta} & e^{-2E_{n-1}\Delta} & \ldots & e^{-(n-1)E_{n-1}\Delta}\\
\end{pmatrix}
\]
with all coefficients distinct and, thus, invertible.
Now, like in Ref.~\cite{Blossier:2009kd}, introduce the dual vectors
$u_k$ with 
\[
(u_k, \chi_l)\ =\ \sum_{i=0}^{n-1} (u_{k}^*)_i \chi_{li}\ =\ \delta_{kl}
\]
for $k,l \in\{0, \ldots, n-1\}$. With these we can write
\begin{equation}
  \begin{split}
    H^0(t)\, u_l\ &=\ \sum_{k=0}^{n-1} c_k e^{-E_k t}\chi_k \chi_k^* u_l\ =\
    c_l e^{-E_l t} \chi_l\\
    &=\ e^{-E_l(t-\tau_0)}\ c_l e^{-E_l \tau_0} \chi_l\\
    &=\ e^{-E_l(t-\tau_0)} H^0(\tau_0)\, u_l\\
  \end{split}
\end{equation}
Thus, the GEVP \eq{eq:hankelgevp} is solved by
\begin{equation}
  \label{eq:hankelsol}
  \Lambda^0_k(t, \tau_0)\ =\ e^{-E_k(t-\tau_0)}\,,\quad
  v_k\ \propto\ u_k\,.
\end{equation}
Moreover, much like in the case of the GEVM
we get the orthogonality 
\begin{equation}
  \label{eq:ortho}
  (u_l,\, H^0(t) u_k)\ =\ c_l e^{-E_l t}\delta_{lk}\,,\quad k,l\in\{0, \ldots,
  n-1\}
\end{equation}
for all $t$-values, because $H^0(t)\, u_k\propto \chi_k$.

\subsubsection{Global PGEVM}

In practice, there are two distinct ways to solve the
GEVP~\eq{eq:hankelgevp}: one can fix $\tau_0$ and determine
$\Lambda^0_k(\tau_0, t)$ 
as a function of $t$. In this case the solution \eq{eq:hankelsol}
indicates that for each $k$ the eigenvalues decay exponentially in
time. On the other hand, one can fix $\delta t = t-\tau_0$ and determine
$\Lambda^0_k(\tau_0, \delta t)$ as a function of $\tau_0$. In this
case the solution \eq{eq:hankelsol} reads
\[
\Lambda^0_k (\tau_0, \delta t)= e^{-E_k\,\delta t}\ =\ \mathrm{const}\,,
\]
because $\delta t$ is fixed.

The latter approach allows to formulate a global PGEVM. Observing that
the matrices $\chi$ do not depend on $\tau_0$, one can reformulate the
GEVP~\eq{eq:hankelgevp} as follows
\begin{equation}
  \label{eq:globalhankel}
  \sum_{\tau_0} H^0(\tau_0+\delta t)\, v_l\ =\ \Lambda^0_l(\delta t)\,
  \sum_{\tau_0} H^0(\tau_0)\, v_l\,, 
\end{equation}
since $\Lambda^0_k$ does not depend on $\tau_0$. However, this works
only as long as there are only $n$ states contributing and all these
$n$ states are resolved by the PGEVM, as will become
clear below. If this is not the case, pollutions and resolved states
will change roles at some intermediate $\tau_0$-value.

\subsubsection{Effects of Additional States}

Next, we ask the question what corrections to the above result we
expect if there are more than $n$ states contributing, i.e.\@
a correction term
\begin{equation}
  C^1(t)\ =\ \sum_{k=n}^{\infty} c_k\, e^{-E_k t}
\end{equation}
to the correlator and a corresponding correction to the Hankel matrix
\[
\epsilon H_{ij}^1(t)\ =\ \sum_{k=n}^\infty C^1(t + i + j)\,.
\]
(We have set $\Delta=1$ for simplicity.)
We assume that we work at large enough $t$ such that these corrections
can be considered as a small perturbation. Then it turns out that the
results of Refs.~\cite{Luscher:1990ck,Blossier:2009kd} apply directly
to the PGEVM and all systematics are identical (\eq{eq:t0corr} or
\eq{eq:t0corr2}). 

However, there is one key difference between GEVM and PGEVM. The GEVM
with periodic boundary conditions is not able to distinguish the
forward and backward propagating terms in
\[
c\left(e^{-E t} \pm e^{-E(T-t)}\right)\,,
\]
as long as they come with the same amplitude. In fact, the eigenvalue
$\lambda^0$ will in this case also be a $\cosh$ or
$\sinh$~\cite{Irges:2006hg}. In contrast, the PGEVM can distinguish
these two terms. As a consequence, the backward propagating part needs
to be treated as a perturbation like excited states and $\Lambda^0$ is
no longer expected to have a $\cosh$ or $\sinh$ functional form in the
presence of periodic boundary conditions.

This might seem to be a disadvantage at first sight. However, we will
see that this does not necessarily need to be the case.

Concerning the size of corrections there are two regimes to
consider~\cite{Blossier:2009kd}: when $\tau_0$ is fixed at small or
moderately large values and $\Lambda$ is studied as a function of
$t\to\infty$ the corrections of the form~\eq{eq:t0corr}
apply~\cite{Luscher:1990ck}.  When, on the other hand, $\tau_0$
is fixed but $\tau_0\geq t/2$ is chosen and the effective
masses~\eq{eq:effmass} of the eigenvalues are studied, corrections
are reduced to $\mathcal{O}(e^{-\Delta E_{n,l}t})$ with
$\Delta E_{m,n} = E_m - E_n$~\cite{Blossier:2009kd}.

$\tau_0\geq t/2$ is certainly fulfilled if we fix $\delta t$ to some
(small) value. However, for this case $\Lambda^0(t, \tau_0)$ is
expected to be independent of both, $t$ and $\tau_0$ when ground state
dominance is reached and $\Meff$ is, thus, not applicable. Therefore,
we define alternative effective masses 
\begin{equation}
  \label{eq:Eeff}
  \tilde{M}_{\mathrm{eff},l}(\delta t,\tau_0)\ =\ -\frac{\log\left(\Lambda_l(\delta t,
  \tau_0)\right)}{\delta t}
\end{equation}
and apply the framework from Ref.~\cite{Blossier:2009kd} to determine
deviations of $\tilde{M}_{\mathrm{eff},l}$ from the true $E_l$. The authors of
Ref.~\cite{Blossier:2009kd} define $\epsilon = e^{-(E_{n}-E_{n-1})\tau_0}$
and expand
\begin{equation}
  \Lambda_l = \Lambda^0_l + \epsilon \Lambda^1_l + \epsilon^2\Lambda^2_l
  + \ldots\,,
\end{equation}
where we denote the eigenvalues of the full problem as $\Lambda(t, \tau_0)$.
Already from here it is clear that in the situation with $\delta t$
fixed and $\tau_0\to\infty$ the expansion parameter $\epsilon$ becomes
arbitrarily small. Simultaneously with $\tau_0$ also $t\to\infty$. The
first order correction (which is dominant for $\tau_0\geq t/2$) to
$\Lambda_l$ reads 
\begin{equation}
  \begin{split}
    \Lambda_l(\delta t, \tau_0) = e^{-E_l\delta t} 
    +& \frac{c_n}{c_l} e^{-(\Delta E_{n,l})\tau_0}\,\\
    \times &\left(e^{-\Delta
        E_{n,l}\delta t}-1\right) c_{l, n}\\
  \end{split}
\end{equation}
with the definition of $\Delta E_{m,n}$ from above and constant
coefficients 
\[
c_{l, n}\ =\ (v_l^0, \chi_{n})(\chi_{n}, v_l^0)\,.
\] 
These corrections are decaying exponentially in $\tau_0$ with a
decay rate determined by $\Delta E_{n,l}$ as expected from
Ref.~\cite{Blossier:2009kd}. For the effective energies we find 
\begin{equation}
  \label{eq:Eeffcorr}
  \begin{split}
    \tilde{M}_{\mathrm{eff},l}(\delta t,\tau_0)\ \approx \ E_l 
    &+\frac{c_n}{c_l}
    e^{-(\Delta E_{n,l})\tau_0}\,\\
    &\times \left(e^{E_l\delta
      t}-e^{-E_{n}\delta t} \right)
    \frac{c_{l, n}}{\delta t}\,,\\
  \end{split}
\end{equation}
likewise with corrections decaying exponentially in $\tau_0$, again
with a rate set by $\Delta E_{n,l}$.

\subsection{Combining GEVM and PGEVM}

There is one straightforward way to combine GEVM and PGEVM: we noted
already above that the principal correlators of the GEVM are again a
sum of exponentials, and, hence, the PGEVM can be applied to them.
This means a sequential application of first the
GEVM with a correlator matrix of size $n_0$ to determine principal
correlators $\lambda_k$ and then of the PGEVM with size $n_1$ and the
$\lambda_k$'s as input. This combination allows us to work with two
relatively small matrices, which might help to stabilise the method
numerically. Moreover, the PGEVM is 
applied only for the respective ground states in the principal
correlators and only relatively small values of $n_1$ are needed. 

An additional advantage lies in the fact that $\lambda_k$ is a sum of
exponentials with only positive coefficients, because it represents a
correlation function with identical operators at source and sink. As a
consequence, the Hankel matrix $H^0$ is positive definite. 

\subsection{Generalised Pencil of Function (GPOF)}

For certain cases, the PGEVM can actually be understood as a special
case of the generalised pencil-of-function (GPOF) method, see
Refs.~\cite{Aubin:2010jc,Aubin:2011zz,Schiel:2015kwa,Ottnad:2017mzd}
and references therein. Making use of the time evolution operator, we
can define a new operator
\begin{equation}
  \hat O_{\Delta t}(t')\ \equiv\ \hat O(t'+ \Delta t) = \exp(H
  \Delta t)\ \hat O(t')\ \exp(-H \Delta t)\,.
\end{equation}
This allows us to write
\begin{equation}
  \langle \hat O_i^{}(t')\ \hat O_j^\dagger(t'+t + \Delta t)\rangle\ =\
  \langle \hat O_i^{}(t')\ \hat O_{\Delta t,j}^\dagger(t'+t)\rangle\,,
\end{equation}
which is the same as $C_{ij}(t + \Delta t)$. Using $i=j$ and the
operators $O_i$, $O_{\Delta t, i}$, $O_{2\Delta t, i}, \dots$ one
defines the PGEVM based on a single correlation function. Note,
however, that the PGEVM is more general as it is also applicable to
sum of exponentials not stemming from a two-point function.

The generalisation is now straightforward by combining $\hat O_i$ and 
$\hat O_{m\Delta t, i}$ for $i=0, \ldots, n_0-1$ and $m=0, \ldots,
n_1-1$. These operators define a Hankel matrix $\mathcal{H}^0$ with
size  $n_1$ of correlator matrices of size $n_0$ as follows
($\Delta=1$ for simplicity) 
\begin{equation}
  \mathcal{H}^0_{\alpha\beta}\ =\ \sum_{k=0}^{n'-1} e^{-E_k t}
  \eta_{k\alpha}\eta_{k\beta}^*\,, 
\end{equation}
with
\begin{equation}
  (\eta_{k})_{i n_0 + j}\ =\ e^{-E_k i}\psi_{kj}\,,
\end{equation}
for $j=0, \ldots,n_0-1$ and $i=0, \ldots, n_1-1$.
Then $n'=n_0\cdot n_1$ is the number of energies that can be resolved.
$\mathcal{H}$ is hermitian, positive definite and the
same derivation as the one from the previous subsection leads to the
GEVP
\[
\mathcal{H}^0(t)\, v_k\ =\ \Lambda_k(t, \tau_0)\, \mathcal{H}^0(\tau_0)\, v_k
\]
with solutions
\[
\Lambda^0_k(t, \tau_0) = e^{-E_k(t - \tau_0)}\,.
\]
In this case the matrix $\mathcal{H}^0$ is positive definite, but
potentially large, which might lead to numerical instabilities. This
can be alleviated by using only for a limited subset of
operators $\hat O_i$ their shifted versions $\hat O_{m\Delta t,i}$,
preferably for those $\hat O_i$ contributing the least noise.

\section{Numerical Implementation}

In case the Hankel matrix $H^0$ is positive definite,
one can compute the Cholesky decomposition $C(t_0)=L\cdot L^T$. Then
one solves the ordinary eigenvalue problem
\[
L^{-1}\, C(t)\, L^{-T}\, w_k = \lambda_k w_k
\]
with $w_k= L^T v_k$.

If this is not the case, the numerical solution of the PGEVM can proceed along
two lines. The first is to compute the inverse of $H^0(\tau_0)$ for
instance using a QR-decomposition and then solve the ordinary
eigenvalue problem for the matrix $A=H^0(\tau_0)^{-1}H^0(t)$. Alternatively, one may take
advantage of the symmetry of both $H^0(t)$ and $H^0(\tau_0)$. One
diagonalises both $H^0(t)$ and $H^0(\tau_0)$ with diagonal eigenvalue
matrices $\Lambda_t$ and $\Lambda_{\tau_0}$ and orthogonal eigenvector matrices
$U_t$ and $U_{\tau_0}$. Then, the eigenvectors of the generalised problem
are given by the matrix
\[
U\ =\ U_{\tau_0}\, \Lambda_{\tau_0}^{-1/2}\, U_t
\]
and the generalised eigenvalues read
\[
\Lambda\ =\ U^T\, H^0(t)\, U\,.
\]
Note that $U$ is in contrast to $U_t$ and $U_{\tau_0}$ not orthogonal.

\subsection{Algorithms for sorting GEVP states}
\label{sec:sorting}

Solving the generalized eigenvalue problem in Eq.~(\ref{eq:GEVP}) for
an $n\times n$ correlation function matrix $C(t)$ (or Hankel matrix
$H$) with $t>t_0$, results in an a priori unsorted set $\l\{s_k(t)
|k\in[0,...,n-1]\r\}$ of states $s_k(t)=(\lambda_k(t,t_0),
\vec{v}_k(t,t_0))$ on each timeslice $t$ defined by an eigenvalue
$\lambda_k(t,t_0)$ and an eigenvector $\vec{v}_k(t,t_0)$. In the
following discussion we assume that the initial order of states is
always fixed on the very first timeslice $t_0+1$ by sorting the states
by eigenvalues, i.e.\ choosing the label $n$ by requiring
$\lambda_0(t_0+1,t_0) > \lambda_1(t_0+1,t_0) > ... >
\lambda_{n-1}(t_0+1,t_0)$, s.t.\ the vector of states reads
$(s_{0}(t_0+1), ..., s_{n-1}(t_0+1) )$. \par  

After defining the initial ordering of states, there are many different possibilities to sort the remaining states for $t>t_0$. In general, this requires a prescription that for any unsorted vector of states $(s_{(k=0)}(t), ..., s_{(k=n-1)}(t) )$ yields a re-ordering $s_{\epsilon(k)}(t)$ of its elements. The permutation $\epsilon(k)$ may depend on some set of reference states $(s_{0}(\tilde{t}), ... , s_{n-1}(\tilde{t}))$ at time $\tilde{t}$ which we assume to be in the desired order. However, for the algorithms discussed here, such explicit dependence on a previously determined ordering at a reference time $\tilde{t}$ is only required for eigenvector-based sorting algorithms. Moreover, $\tilde{t}$ does not necessarily have to equal $t_0+1$. In fact, the algorithms discussed below are in practice often more stable for choosing e.g.\ the previous timeslice $t-1$ to determine the order of states at $t$ while moving through the available set of timeslices in increasing order. \par

\subsubsection{Sorting by eigenvalues}
This is arguably the most basic way of sorting states; it simply consists of repeating the ordering by eigenvalues that is done at $t_0$ for all other values of $t$, i.e.\ one chooses $\epsilon(k)$ independent of any reference state and ignoring any information encoded in the eigenvectors, s.t.\@
\begin{equation}
 \lambda_0(t,t_0) > \lambda_1(t,t_0) > ... > \lambda_{n-1}(t,t_0) \,.
 \label{eq:eval_sort}
\end{equation}
The obvious advantage of this method is that it is computationally fast and trivial to implement. However, it is not stable under noise which can lead to a rather large bias and errors in the large-$t$ tail of the correlator due to incorrect tracking of states. This is an issue for systems with a strong exponential signal-to-noise problem (e.g.\ the $\eta$,$\eta'$-system) as well as for large system sizes $n$. Moreover, the algorithm fails by design to correctly track crossing states, which causes a flipping of states at least in an unsupervised setup and tends to give large point errors around their crossing point in $t$. \par

\subsubsection{Simple sorting by eigenvectors}
Sorting algorithms relying on eigenvectors instead of eigenvalues generally make use of orthogonality properties. A simple method is based on computing the scalar product
\begin{equation}
 c_{kl} = \langle \vec{v}_l(\tilde{t}), \vec{v}_k(t)\rangle \,,
 \label{eq:evec_simple_sort}
\end{equation}
where $\vec{v}_l(\tilde{t})$ denote eigenvectors of some (sorted) reference states $s_l(\tilde{t})$ at $\tilde{t}<t$ and $\vec{v}_k(t)$ belongs to a state $s_k(t)$ that is part of the set which is to be sorted. For all values of $k$ one assigns $k\rightarrow \epsilon(k)$, s.t.\@ $\l| c_{kl} \r| \stackrel{!}{=} \mathrm{max}$. If the resulting map $\epsilon(k)$ is a permutation the state indexing at $t$ is assigned according to $s_{k}(t) \rightarrow s_{\epsilon(k)}(t)$. Otherwise sorting by eigenvalues is used as a fallback. \par

This method has some advantages over eigenvalue-based sorting methods: It can in principle track crossing states and flipping or mixing of states in the presence of noise are less likely to occur. The latter is especially an issue for resampling (e.g.\ bootstrap or jackknife), i.e.\ if state assignment fails only on a subset of samples for some value(s) of $t$, leading to large point errors and potentially introducing a bias. On the downside, the resulting order of states from this method is in general not unambiguous for systems with $n>2$ and the algorithm is not even guaranteed to yield a valid permutation $\epsilon(k)$ for such systems in the presence of noise, hence requiring a fallback.\par

\subsubsection{Exact sorting by eigenvectors}
\label{subec:exact_sorting}
Any of the shortcomings of the aforementioned methods are readily avoided by an approach that uses volume elements instead of scalar products. This allows to obtain an unambiguous state assignment based on (globally) maximized orthogonality. The idea is to consider the set of all possible permutations $\{\epsilon(k)\}$ for a given $n\times n$ problem and compute 
\begin{equation}
  \label{eq:evec_exact_sort}
  \begin{split}
    c_{\epsilon} =\prod_k & \l| \mathrm{det}\l(
    \vec{v}_{0}(\tilde{t}), \ldots , \vec{v}_{\epsilon(k)-1}(\tilde{t})\right.\right.
    ,\\ &\left.\left.\vec{v}_k(t), \vec{v}_{\epsilon(k)+1}(\tilde{t}) , \ldots , \vec{v}_{n-1}(\tilde{t})\r) \r| \,,\\
  \end{split}
\end{equation}
for each $\epsilon$. This can be understood as assigning a score for how well each individual vector $v_k(t)$ fits into the set of vectors at the reference timeslice $\tilde{t}$ at a chosen position $\epsilon(k)$ and computing a global score for the current permutation $\epsilon$ by taking the product of the individual scores for all vectors $v_k(t)$. The final permutation is then chosen s.t.\ $c_{\epsilon}\stackrel{!}{=}\mathrm{max}$. \par

Unlike the method using the scalar product, this method is guaranteed
to always give a unique solution, which is optimal in the sense that
it tests all possible permutations and picks the global
optimum. Therefore, the algorithm is most stable under noise and well
suited for systems with crossing states. Empirically, this results in
e.g.\ the smallest bootstrap bias at larger values of $t$ compared to
any other method described here. A minor drawback of the approach is
that it is numerically more expensive due to the required evaluations
of (products of) volume elements instead of simple scalar
products. However, this becomes only an issue for large system sizes
and a large number of bootstrap (jackknife) samples.

\subsubsection{Sorting by minimal distance}
\label{subsec:mindistsort}
While the methods discussed above work all fine for the standard case
where the GEVP is solved with fixed time $t_0$ (or $\tau_0$) and
$\delta t$ is varied, the situation is different for $\tMeff$ with
$\delta t$ fixed: there are $t$-values for which it is numerically not
easy to separate wanted states from pollutions, because they are of
very similar size in the elements of the sum of exponentials entering
at these specific $t$-values. However, when looking at the bootstrap
histogram of all eigenvalues, there is usually a quite clear peak at
the expected energy value for all $t$-values with not too much noise.

Therefore, we implemented an alternative sorting for this situation
which goes by specifying a target value $\xi$. Then we chose among all
eigenvalues for a bootstrap replicate the one which is closest to
$\xi$. The error is computed from half of the $16$\% to $84$\%
quantile distance of the bootstrap distribution and the central value
as the mean of $16$\% and $84$\% quantiles. For the central value one
could also use the median, however, we made the above choice to have
symmetric errors.

This procedure is much less susceptible to large outliers in the
bootstrap distribution, which appear because of the problem discussed
at the beginning of this sub-section.

For the numerical experiments shown below we found little to no
difference in between sorting by \emph{eigenvalues} and
any of the sorting by \emph{vectors}. Thus, we will work with sorting by
\emph{eigenvalues} for all cases where we study $\Lambda_l(t, \tau_0)$ with
$\tau_0$ fixed. On the other hand, specifying a target value $\xi$
and sort by \emph{minimal distance}
turns out to be very useful for the case $\Lambda_l(\delta t, \tau_0)$
with $\delta t$ fixed. As it works much more reliably than the other
two approaches, we use this sorting by \emph{minimal distance} for the
$\delta t$ fixed case throughout this paper.

The methods used in this paper are fully implemented in a R package
called \emph{hadron}~\cite{hadron:2020}, which is freely available
software.

\section{Numerical Experiments}
\label{sec:experiments}

\begin{figure*}[th]
  \centering
  \subfigure{\includegraphics[width=.4\textwidth,page=1]{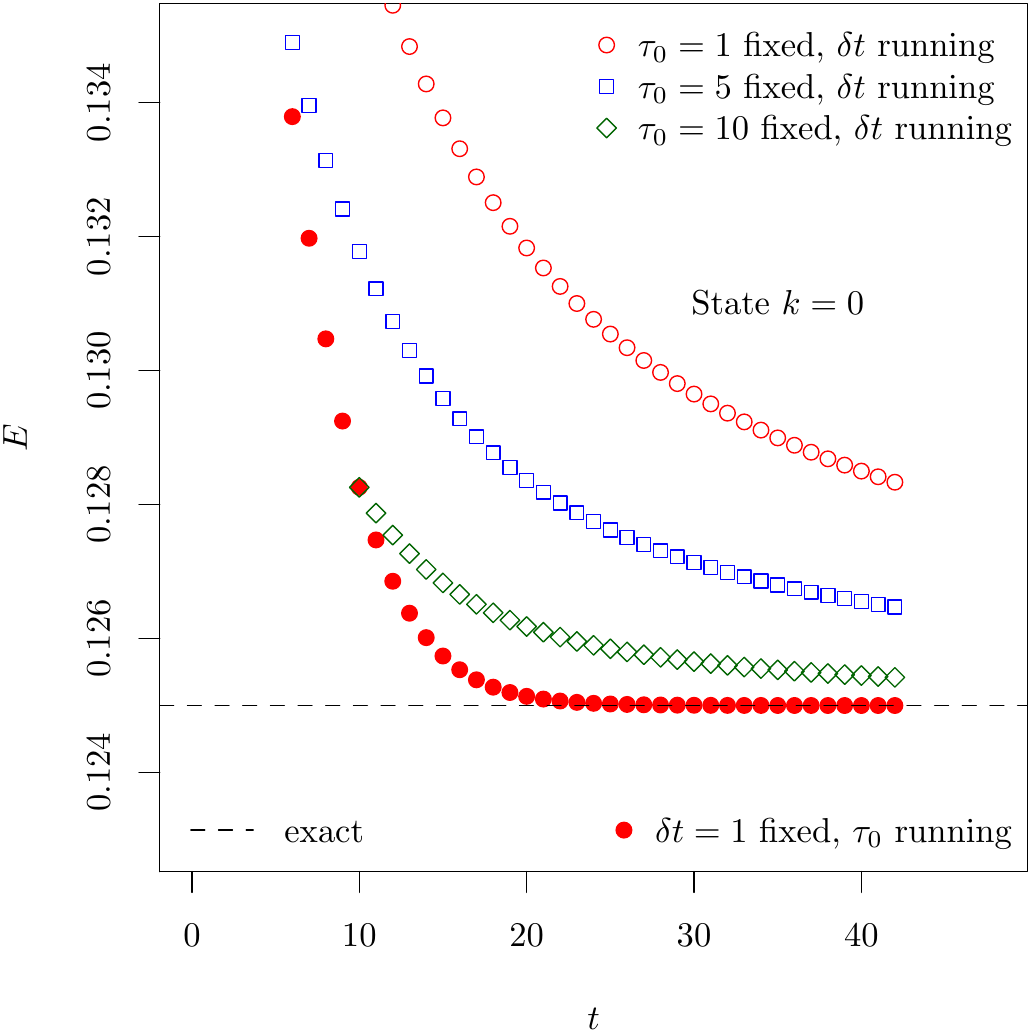}}\qquad
  \subfigure{\includegraphics[width=.4\textwidth,page=2]{experiments}}
  \caption{Effective energies from the PGEVM with $n=2$ applied
  to synthetic data containing three states with $\Delta=1$. Open symbols correspond
  to $\tMeff$ \eq{eq:Eeff} of the Prony principal
  correlator with $\tau_0$
  fixed, while filled symbols are for $\delta t$ fixed. In the
  left panel we show the ground state with $k=0$, in the right one the
  first excited state, both for different choices of $\tau_0$.}
  \label{fig:synthetic}
\end{figure*}

In this section we first apply the PGEVM to synthetic data. With
this we investigate whether additional states not accounted for by the
size of the Prony GEVP lead to the expected distortions in the
principal correlators and effective masses. At this stage the energy
levels and amplitudes are not necessarily chosen realistically, because
we would first like to understand the systematics.

In a next step we apply the combination of GEVM and PGEVM to
correlator matrices from lattice QCD simulations. After applying the
framework to the pion, we have chosen two realistic examples, the
$\eta$-meson and the $\rho$-meson. 

\subsection{Synthetic Data}

As a first test we apply the PGEVM alone to synthetic data. We
generate a correlator
\begin{equation}
C_s(t)\ =\ \sum_{k=0}^{2} c_k\, e^{-E_k t}
\end{equation}
containing three states with $E_k=(0.125, 0.3, 0.5), k=0,1,2$ and
$t\in\{0, \ldots, 48\}$. The amplitudes $c_k$ have been chosen all equal to
$1$. 

\begin{figure}
  \centering
  \includegraphics[width=.4\textwidth,page=3]{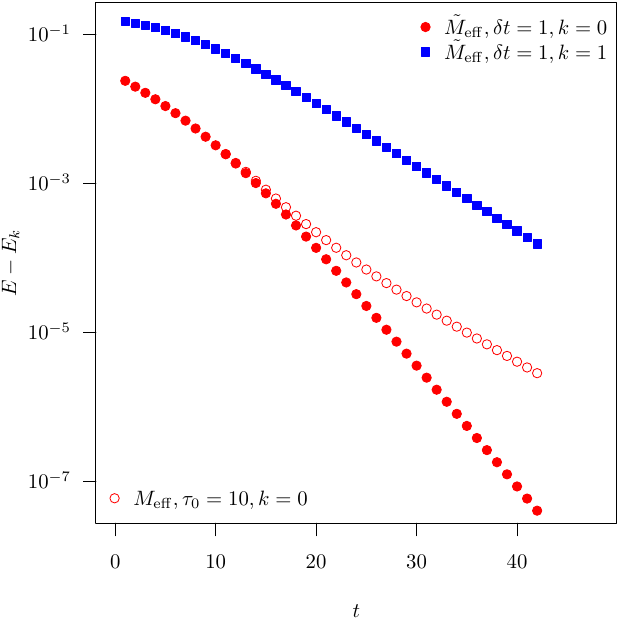}
  \caption{Effective $E$ minus the exact ground
    state energy $E_0=0.125$ for $k=0$ and $E_1=0.3$ for $k=1$ as a
    function of $t$ on a logarithmic scale for $\Delta=1$. Filled symbols correspond
    to $\tMeff$ \eq{eq:Eeff} with $\delta t=1$ fixed, open symbols to
    $\Meff$ \eq{eq:effmass} for $\tau_0=10$ fixed.}
  \label{fig:synthetic4}
\end{figure}

We apply the PGEVM to this correlator $C_s$ with
$n=2$. This allows us to resolve only two states and we would like to see
how much the third state affects the two extracted states. The result
is plotted in \fig{fig:synthetic}. We plot $\tMeff$ of \eq{eq:Eeff} as
a function of $t$, filled symbols correspond to $\delta t=1$
fixed. Open symbols correspond to $\tau_0$ fixed with values
$\tau_0=1, 5$ and $\tau_0=10$. In the left panel we show the ground
state $k=0$, in the right one the second state $k=1$ resolved by the
PGEVM.  The solid lines represent the input values for $E_0$ and
$E_1$, respectively.  

One observes that the third state not resolved by the PGEVM leads
to pollutions at small values of $t$. These pollutions are clearly
larger for the case of fixed $\tau_0$, as expected from our discussion
in section~\ref{sec:method}. The relative size of the pollutions is
much larger in the second state with $k=1$ than in the state with
$k=0$, which is also in line with the expected pollution.

\begin{figure}
  \centering
  \includegraphics[width=.4\textwidth,page=4]{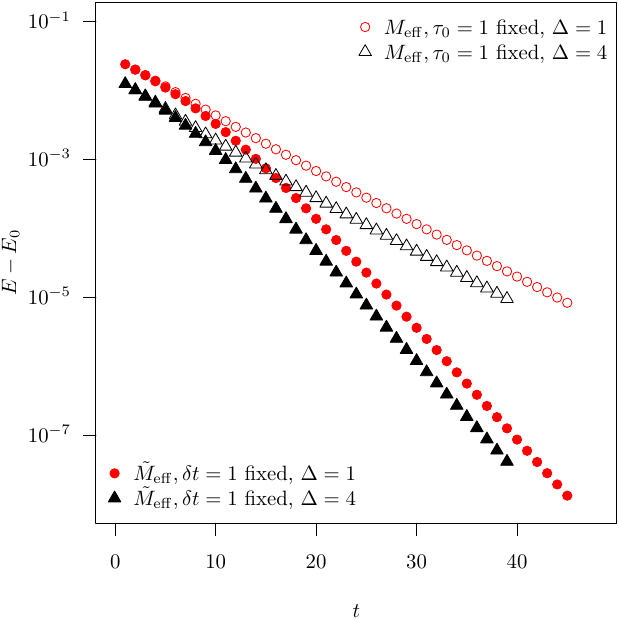}
  \caption{$E-E_0$ for $k=0$ on a logarithmic scale as a function of
    $t$ for different values of $\Delta$. Filled symbols correspond to
    $\tMeff$ with $\delta t=1$ fixed, open symbols to $\Meff$ for $\tau_0=1$ fixed.}
  \label{fig:synthetic5}
\end{figure}

We remark in passing that the not shown values for
$\Meff$ of \eq{eq:effmass} of 
the eigenvalue $\Lambda_k(t, \tau_0)$ at fixed $\tau_0$ are almost
indistinguishable on the scale of \fig{fig:synthetic} from $\tMeff$
with $\delta t$ fixed. For the tiny differences and the influence of
$\tau_0$ thereon see Figures~\ref{fig:synthetic4} and~\ref{fig:synthetic5}.

In \eq{eq:Eeffcorr} we have discussed
that we expect corrections in $\tMeff$ and $\Meff$ to decay
exponentially in $t = \delta t + \tau_0$. We can test this by
subtracting the exactly known energy $E_k$ from the PGEVM
results. Therefore, we plot in \fig{fig:synthetic4} effective masses
minus the exact $E_k$ values as a function of $t$. Filled symbols
correspond to $\tMeff$ with $\delta t=1$ and open symbols (only $k=0$)
to $\Meff$ with $\tau_0=10$. The asymptotically exponential
convergence in $t$ is nicely visible for both effective mass
definitions and also for $k=0$ and $k=1$. For $\tMeff$ the
decay rate is to a good approximation $E_2-E_0$ for $k=0$ and $E_2-E_1$ for
$k=1$, respectively, as expected from \eq{eq:Eeffcorr}. For $\Meff$
the asymptotic logarithmic decay rate is approximately $E_1-E_0$ and,
thus, worse as expected from \eq{eq:t0corr}.

So far we have worked solely with $\Delta=1$. In
\fig{fig:synthetic5} we investigate the dependence of $\tMeff$ and
$\Meff$ on $\Delta$: we plot $E-E_0$ on a logarithmic scale as a
function of $t$ for $\Delta=1$ and $\Delta=4$. While $\Delta$ has no
influence on the convergence rate, it reduces the amplitude of the
pollution for both $\tMeff$ and $\Meff$ by shifting the data points to the 
left.  The reason is that a larger $\Delta$ allows to reach larger times in the 
Hankel matrices at the same $t$. A smaller $\Delta$ on the other hand allows to 
go to larger $t$, thus the advantage of increased $\Delta$ is negligible.

\begin{figure}
  \centering
  \includegraphics[width=.4\textwidth,page=5]{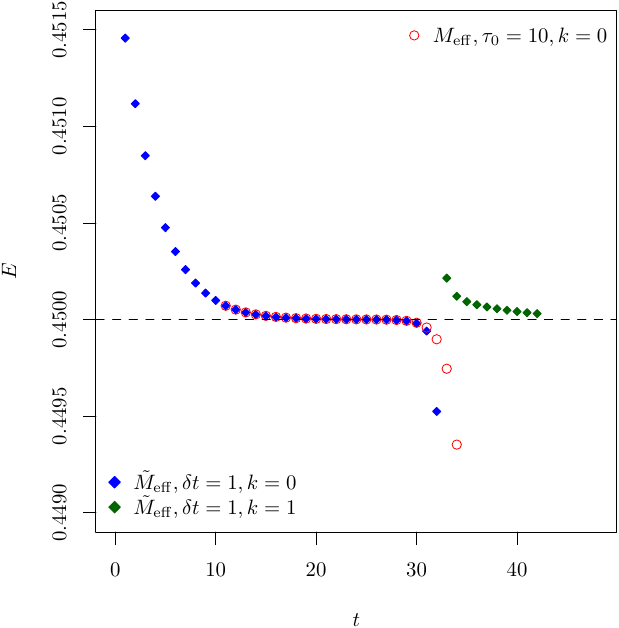}
  \caption{Effective energies as a function of $t$
    for synthetic data including a backpropagating state for the
    $k=0$ ground state obtained by the PGEVM with $\Delta=1$. Filled symbols
    correspond to $\tMeff$ with $\delta t=1$ fixed and open symbols to
    $\Meff$ with $\tau_0=10$ fixed.}
  \label{fig:synthetic2}
\end{figure}

In order to see the effect of so-called back-propagating states,
we next investigate a correlator 
\begin{equation}
C_s(t)\ =\ \sum_{k=0}^{2} c_k\, \left(e^{-E_k t}+ \delta_{k0}e^{-E_k
(T-t)}\right) 
\end{equation}
with a back-propagating contribution to the ground state
$E_0$ only. Energies are chosen as $E_i=(0.45,0.6,0.8)$ and the
amplitudes are $c_i=(1, 0.1, 0.01)$ with $T=96$. The
result for the ground state effective energy determined from the PGEVM
principal correlator is shown in \fig{fig:synthetic2}. We show $\Meff$
from the principal correlator for $\tau_0=10$ fixed as open red symbols. The
filled symbols correspond to $\tMeff$ for $k=0$ and $k=1$ with $\delta
t=1$ fixed. Both is again for $\Delta=1$.

One observes a downward bending of the two $k=0$ effective masses
starting around $t=28$. The difference between $\tau_0$ fixed and
$\delta t$ fixed is only visible in the $t$-range where the bending
becomes significant. Obviously, in this region the contribution of the
forward and backward propagating states becomes comparable in size,
while the state with $k=2$ becomes negligible. Interestingly, for
$\delta t=1$ fixed the state of interest is then contained in the $k=1$
state while the $k=0$ states drop to the state with energy $-E_0$ (not
visible in the figure). 

It becomes clear that there is an intermediate region in $t$, in
this case from $t=28$ to $t=38$, where the different contributions to
the correlator cannot be clearly distinguished by the PGEVM using
$\tMeff$. Around $t=28$ contributions by the $k=2$ state have become
negligible, while the backward propagating state becomes important. At
this point the state with $k=1$ becomes the pollution and the PGEVM
resolves forward and backward propagating states. This transition will
also be visible for the lattice QCD examples discussed next.

\subsection{Lattice QCD Examples}

As a first lattice QCD example 
we start with the charged pion, which gives rise to
one of the cleanest signals in any correlation function extracted from
lattice QCD simulations. In particular, the signal to noise ratio is
independent of $t$. From now on quantities are given in units of the
lattice spacing $a$, i.e.\ $aE$, $aM$, $t/a$, \ldots are dimensionless
real numbers. However, for simplicity we set $a=1$.

The example we consider is the B55.32 ensemble generated with
$N_f=2+1+1$ dynamical quark flavours by ETMC~\cite{Baron:2010bv} at a
pion mass of about $350\ \mathrm{MeV}$. For details on the ensemble we
refer to Ref.~\cite{Baron:2010bv}. The correlation functions for the
pion have been computed with the so-called one-end-trick and spin
dilution, see Ref.~\cite{Boucaud:2008xu} on $4996$ gauge
configurations. The time extent is $T=64$ lattice points, the spatial
one $L=T/2$.

\begin{figure}[th]
  \centering
  \includegraphics[width=.4\textwidth,page=1]{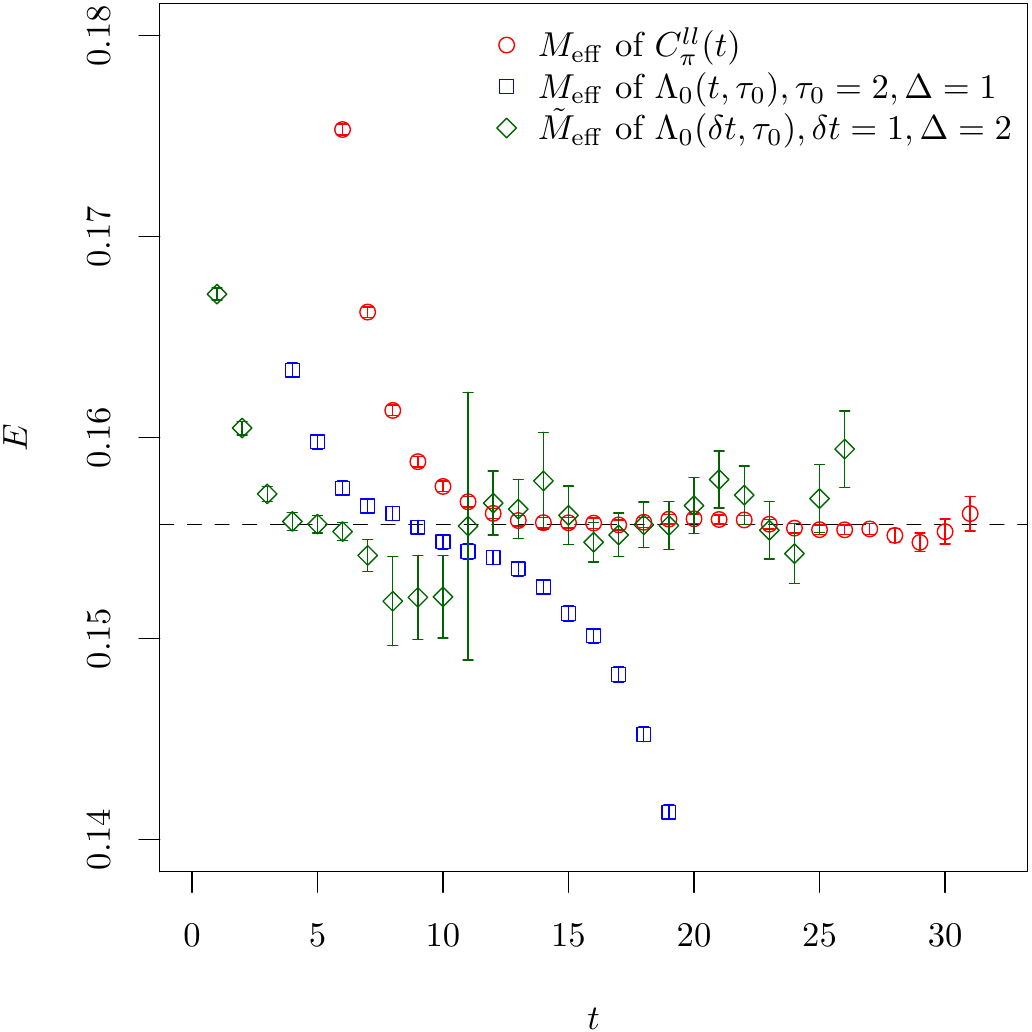}
  \caption{Effective energies $aE$ as a function of $t$
    computed from the local-local two-point 
    pion correlation function on ensemble B55.32. The red circles
    represent the $\cosh$ effective masses \eq{eq:cosheffmass} computed from the
    single twopoint correlator directly. The blue squares are the effective masses
    $\Meff$ computed from the PGEVM principal correlator with
    $\tau_0=2$ and $n=2$
    fixed. The green diamonds represent $\tMeff$
    computed from the PGEVM principal correlator with $\delta t=1$ and $n=2$
    fixed. The dashed line represents the mean value of a fit with a
    two parameter $\cosh$ model to the
    original correlator.}
  \label{fig:pion1}
\end{figure}

\subsubsection{Pion}

We look at the single pion two-point correlation function
$C_\pi^{ll}(t)$ computed with local sink and local source using the
standard operator $\bar u\, i\gamma_5 d$ projected to zero
momentum. Since the pion is relatively light, 
the backpropagating state due to periodic boundary conditions is
important. For this reason, we compute the cosh effective
mass from the ratio
\begin{equation}
  \label{eq:cosheffmass}
  \frac{C_\pi^{ll}(t+1)}{C_\pi^{ll}(t)} = \frac{e^{-E_\pi (t+1)} +
    e^{-E_\pi(T-(t+1))}}{e^{-E_\pi t} + e^{-E_\pi(T-t)}} 
\end{equation}
by solving numerically for $E_\pi$. The corresponding result is shown
as red circles in \fig{fig:pion1} as a function of $t$. The
effective masses $\Meff$ computed from the PGEVM principal
correlator with $\tau_0=2$, $n=2$ and $\Delta=1$ fixed are shown as
blue squares. One observes that excited states are reduced but the
pollution by the backward propagating state ruins the plateau. As
green diamonds we show the $\tMeff$ for the principal correlator with
$\delta t=1$, $n=2$ and $\Delta=2$ fixed. Here, we used a target
value $\xi=0.16$ to identify the appropriate state during
resampling, see section~\ref{sec:sorting}. The plateau
starts as early as $t=5$, there is an intermediate region where
forward and backward propagating states contribute similarly, and
there is a region for large $t$, where again the ground state is
identified. The apparent jump in the data at $t=11$ is related to
coupling to a different state than on previous timeslices and is
accompanied by a large error because the sorting of states is
performed for each bootstrap sample. Coupling to a different state is
allowed for the method with fixed $\delta t$ as the $\tau_0$ of the
GEVP changes for every timeslice. In fact, this feature is a key
difference to the methods with fixed $\tau_0$ for which the set of
states is unambigously determined by the initial choice of $\tau_0$,
see the discussion in section~\ref{subsec:mindistsort}.

\begin{figure}[th]
  \centering
  \includegraphics[width=.4\textwidth,page=2]{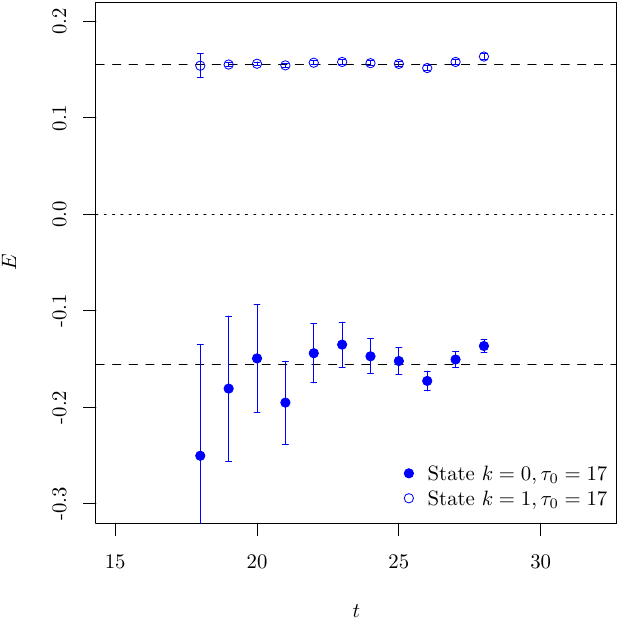}
  \caption{Effective energies $aE$ as a function of $t$
    computed from the local-local twopoint 
    pion correlation function on ensemble B55.32. We show effective
    masses $\Meff$ of the PGEVM principal correlators with $k=0$ and $k=1$ and
    $\tau_0=17$ and $n=2$ fixed.} 
  \label{fig:pion2}
\end{figure}

Once all the excited states have become negligible, the PGEVM can also
resolve both forward and backward propagating states (see also
Ref.~\cite{Schiel:2015kwa}). For the example 
at hand this is shown in \fig{fig:pion2} with $\tau_0=17$ and
$n=2$ fixed. For this to work it is important to chose $\tau_0$ large
enough, such that excited states have decayed
sufficiently. Interestingly, the noise is mainly projected into the
state with negative energy. 

\begin{figure}[th]
  \centering
  \includegraphics[width=.4\textwidth,page=3]{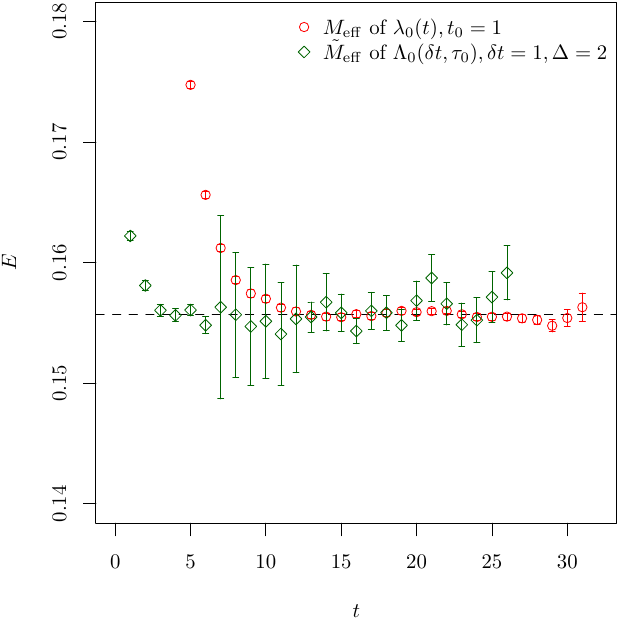}
  \caption{Like \fig{fig:pion1}, but starting with a GEVM
    principal correlator.} 
  \label{fig:pion3}
\end{figure}

In \fig{fig:pion3} we visualise the improvement realised by
combining GEVM with PGEVM. Starting with a $2\times 2$ correlator
matrix built from local and fuzzed operators, we determine the GEVM
principal correlator $\lambda_0(t)$ using $t_0=1$. The $\cosh$ effective
mass of $\lambda_0$ is shown as red circles in
\fig{fig:pion3}. In green we show $\tMeff$
of the PGEVM principal correlator $\Lambda_0$ obtained
with $\delta t= 1$, $n_1=2$ and $\Delta =2$ fixed.

\begin{figure}[th]
  \centering
  \includegraphics[width=.4\textwidth,page=4]{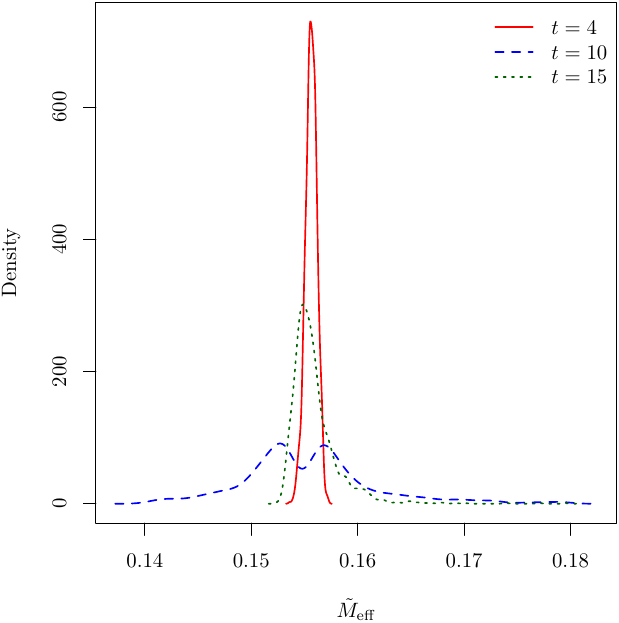}
  \caption{Density of bootstrap replicates for $\tMeff$ at different
    $t$-values for the data of \fig{fig:pion3}}
  \label{fig:pion4}
\end{figure}

Compared to \fig{fig:pion3}, the plateau in $\tMeff$
starts as early as $t=3$. However, in particular at larger
$t$-values the noise is also increased compared to the PGEVM directly
applied to the original correlator. It should be clear that the pion
is not the target system for an analysis combining GEVM and PGEVM,
because its energy levels can be extracted without much systematic
uncertainty directly from the original correlator. However, it serves
as a useful benchmark system, where one can also easily check for
correctness. 

In \fig{fig:pion4} we plot the (interpolated) bootstrap sample
densities of $\tMeff$ for three $t$-values: $t=4$, $t=10$ and
$t=15$. They correspond to the green diamonds in \fig{fig:pion3}. One
observes that at $t=4$ the distribution is approximately
Gaussian. At $t=15$ the situation is similar, just that the
distribution is a bit skew towards larger $\tMeff$-values. In the
intermediate region with $t=10$ there is a two peak structure visible,
which is responsible for the large error. It is explained -- see above
-- by the inability of the method with $\delta t=1$ to distinguish the
different exponentials contributing to $\lambda_0$. 

\begin{table}
  \centering
  \begin{tabular*}{.97\linewidth}{@{\extracolsep{\fill}}lrrrr}
    \hline
    & $t_1$ & $t_2$ & $\Delta$ & $M_\pi$\\
    \hline
    $M_\mathrm{eff}$ of $C_\pi^{ll}$ &15 & 30 & - & $0.15567(12)$\\
    $\tMeff$ of PGEVM & 4 & 20 & 1 & $0.15539(25)$\\
    $\tMeff$ of GEVM/PGEVM & 3 & 20 & 2 & $0.15569(25)$\\
    \hline
  \end{tabular*}
  \caption{Results for $M_\pi$ of fits to various pion effective
    energies, see red circles and green diamonds of \fig{fig:pion1}
    for the first two rows and green diamonds of \fig{fig:pion2} for
    the third row. The fit ranges are $[t_1, t_2]$.}
  \label{tab:pionres}
\end{table}

In \tab{tab:pionres} we have compiled fit results obtained for the
pion: the first row corresponds to a fit to the effective mass of the
correlator $C_\pi^{ll}$ in the fit range indicated by $t_1, t_2$. The
second row represents the fit to $\tMeff$ with $\delta t=1$
fixed obtained with PGEVM on $C_\pi^{ll}$ directly (green diamonds in
\fig{fig:pion1}). The last row is the same, but for the combination of
GEVM/PGEVM (green diamonds in \fig{fig:pion3}).
The agreement is very good, even though the PGEVM and GEVM/PGEVM
errors are larger than the ones obtained from the correlator directly.

\subsubsection{$\eta$-meson}

\begin{figure}[th]
  \centering
  \includegraphics[width=.4\textwidth,page=1]{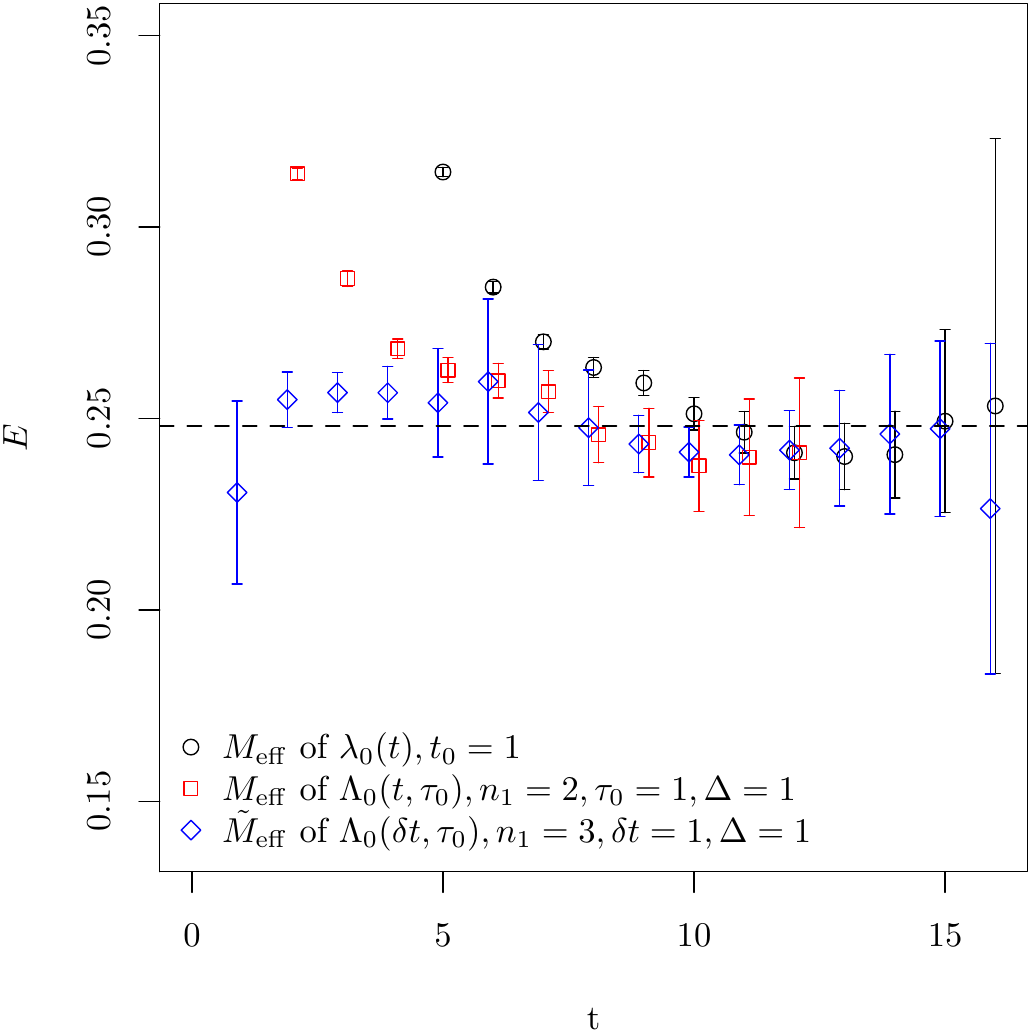}
  \caption{Effective energies for the $\eta$-meson as a function of $t$ for
  the GEVM principal correlator $\lambda_0$ and from
  the GEVM/PGEVM principal correlator $\Lambda_0$ with $n_1=2$ and
  $n_1=3$, respectively. The data is for the B55.32 ETMC ensemble. The
  dashed horizontal line represents the result quoted in
  Ref.~\cite{Ottnad:2017bjt}.} 
  \label{fig:eta}
\end{figure}

As a next example we study the $\eta/\eta^\prime$ system,
where due to mixing of flavour singlet and octet states the
GEVM cannot be avoided in the first place. In addition, due to
large contributions by fermionic disconnected diagrams the correlators
are noisy making the extraction of energy levels at late Euclidean
times difficult. The $\eta/\eta^\prime$ analysis on the B55.32
ensemble was first carried out
in Refs.~\cite{Ottnad:2012fv,Michael:2013gka,Ottnad:2017bjt} using a
powerful method to subtract excited states we can compare to. However,
this excited state subtraction method is based on some (well founded)
assumptions. 

The starting point is a $3\times 3$ correlator matrix $C_{ij}^\eta(t)$
with light, strange and charm flavour singlet operators and local
operators only. We apply the GEVM with $t_0=1$ and extract the first
principal correlator $\lambda_0(t)$ corresponding to the $\eta$-state,
which is then input to the PGEVM. 

In \fig{fig:eta} we show the effective mass of the $\eta$-meson for
this GEVM principal correlator $\lambda_0(t)$ as black circles. 
In addition we show as red squares the effective masses of $\Lambda_0$
obtained from the PGEVM applied to this principal correlator with
$n_1=2$, $\tau_0=1$ and $\Delta=1$. The blue diamonds represent $\tMeff$
of $\Lambda_0$ obtained with $n_1=3$, $\delta t=1$ and $\Delta =1$
fixed. The dashed horizontal line indicates the results obtained 
using excited state subtraction~\cite{Ottnad:2017bjt}. For better
legibility we show the effective masses for each of the three cases
only up to a certain $t_\mathrm{max}$ after which errors become too
large. Moreover, the two PGEVM results are slightly displaced horizontally.

One observes two things: excited state pollutions are
significantly reduced by the application of the PGEVM to
the GEVM principal correlator $\lambda_0$. However, also noise
increases. But, since in the effective masses of $\lambda_0$ there are
only $5$ points which can be interpreted as a plateau, the usage of
PGEVM significantly increases the confidence in the analysis. 

In the corresponding $\eta^\prime$ principal correlator the noise is
too large to be able to identify a plateau for any of the cases
studied for the $\eta$. 

\begin{table}
  \begin{tabular*}{.9\linewidth}{@{\extracolsep{\fill}}lrrrr}
    \hline
    & $t_1$ & $t_2$ & $\Delta$ & $M_\eta$\\
    \hline
    $M_\mathrm{eff}$ of $\lambda_0$ & 10 & 16 & - & $0.2467(29)$\\
    $M_\mathrm{eff}$ of $\Lambda_0$, $t_0=1$ & 7 & 14 & 1 & $0.2425(38)$\\
    $\tilde{M}_\mathrm{eff}$ of $\Lambda_0$, $\delta t=1$ & 1 & 15 & 1
    & $0.2504(36)$\\
    \hline
    Ref.~\cite{Ottnad:2017bjt} & - & - & - & $0.2481(08)$\\
    \hline
  \end{tabular*}
  \caption{Results of fits to effective $\eta$ energies, see
    \fig{fig:eta}. The fitrange is given by $[t_1, t_2]$.}
  \label{tab:etares}
\end{table}

In table \tab{tab:etares} we present fit results to the different
$\eta$ effective masses from \fig{fig:eta}. The agreement among the
different definitions, but also with the literature value is
reasonable within errors.

\subsubsection{$I=1, \pi-\pi$-scattering} 

\begin{figure*}[th]
  \centering
  \subfigure{\includegraphics[width=.4\textwidth,page=2]{rho}}\qquad
  \subfigure{\includegraphics[width=.4\textwidth,page=1]{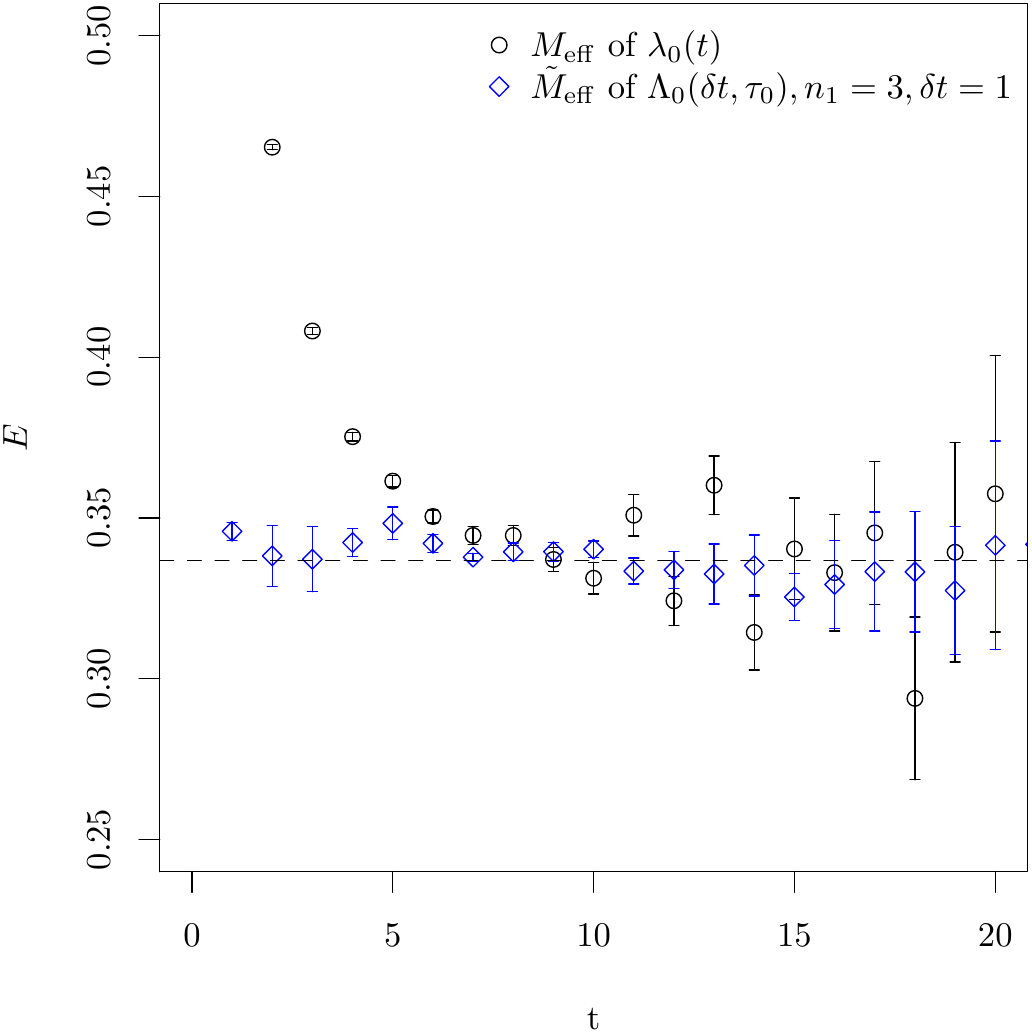}}
  \caption{Effective energies for $I=1, \pi-\pi$-scattering as a function of
    $t$. Left: $A_1$ irrep with total momentum square equal to $1$
    in lattice units. Right: $T_{1u}$ irrep with total zero
    momentum. In both cases the ground state energy level is
    shown. The dashed horizontal lines represent the fit results to
    $\Meff$ of $\lambda_0$, see Tables~\ref{tab:rhores2} and
    \ref{tab:rhores1}.} 
  \label{fig:rho}
\end{figure*}

Finally, we investigate correlator matrices for the $I=1,
\pi-\pi$-scattering. The 
corresponding correlator matrices were determined as part of a Lüscher
analysis including moving frames and all relevant lattice irreducible
representations (irreps). A detailed discussion of the framework and the theory can be
found in Ref.~\cite{Werner:2019hxc}. Here we use the
$N_f=2$ flavour ensemble cA2.30.48 generated by
ETMC~\cite{Abdel-Rehim:2015pwa,Liu:2016cba}, to which we apply the
same methodology as discussed in Ref.~\cite{Werner:2019hxc}.

The first example corresponds to the ground state in the $A_1$
irreducible representation with total squared momentum equal to $1$ in
units of $4\pi^2/L^2$, for which the results are shown in the left
panel of \fig{fig:rho}. In this case the effective mass computed from
the GEVM principal correlator $\lambda_0$ shows a reasonable plateau
(black circles). The red squares show $\Meff$ of
$\Lambda_0$ with $n_1=2$, $\tau_0=1$ and $\Delta=2$ fixed. Even though
the plateau starts at earlier times, noise is increasing
quickly. Actually, we no longer display the energies from $t>17$ due to
too large error bars for better legibility. When using $\tMeff$ with
$n_1=3$, $\delta t=1$ and $\Delta=1$, a plateau can be
identified from $t=1$ on and with a very reasonable signal to noise
ratio. 

\begin{table}
  \centering
  \begin{tabular*}{.9\linewidth}{@{\extracolsep{\fill}}lrrr}
    \hline
    & $t_1$ & $t_2$ & $aW_\rho$\\
    \hline
    $M_\mathrm{eff}$ of $\lambda_0$ & 9 & 20 & $0.28411(26)$\\
    $M_\mathrm{eff}$ of $\Lambda_0$ with $t_0=3$ & 2 & 15 & $0.28235(28)$\\
    $\tilde{M}_\mathrm{eff}$ of $\Lambda_0$ with $\delta t=1$ & 1 & 20 & $0.2838(10)$\\
    \hline
  \end{tabular*}
  \caption{Results of fits to effective energy levels
    for $I=1, \pi-\pi$-scattering for the $A_1$ 
    irrep, see left panel of \fig{fig:rho}.}
  \label{tab:rhores2}
\end{table}

Fit results to the effective masses for the $A_1$ irrep are compiled
in \tab{tab:rhores2}. Here one notices that, despite the visually much
longer plateau range, the error on the fitted mass is significantly
larger for $\tMeff$ than for the other two methods. The overall
agreement is very good, though.

The same can be observed in the right panel of \fig{fig:rho} for the
$T_{1u}$ irrep. However,
this time it is not straightforward to identify a plateau in $\Meff$
of $\lambda_0$ shown as black circles. Using $\tMeff$ instead
with $n_1=3$, $\delta t=1$ and $\Delta = 1$ fixed improves
significantly over the traditional effective masses and give much
higher confidence to the extracted energy levels.

\begin{table}
  \centering
  \begin{tabular*}{.9\linewidth}{@{\extracolsep{\fill}}lrrr}
    \hline
    & $t_1$ & $t_2$ & $aW_\rho$\\
    \hline
    $M_\mathrm{eff}$ of $\lambda_0$ & 9 & 20 & $0.33680(67)$\\
    $\tilde{M}_\mathrm{eff}$ of $\Lambda_0$ with $\delta t=1$ & 2 & 20 & $0.3377(16)$\\
    \hline
  \end{tabular*}
  \caption{Results of fits to effective energy levels for $I=1,
    \pi-\pi$-scattering for the $T_{1u}$ 
    irrep, see right panel of \fig{fig:rho}.}
  \label{tab:rhores1}
\end{table}

Fit results for the $T_{1u}$ irrep are compiled in
\tab{tab:rhores1}. The conclusion is similar to the one from the $A_1$
irrep.

\section{Discussion}

In this paper we have first discussed the relation among the
generalised eigenvalue, the Prony and the generalised pencil of
function methods: they are all special cases of a generalised
eigenvalue method. This fact allows one to discuss systematic effects
stemming from finite matrix sizes used to resolve the infinite tower
of states. The results previously derived for the generalised
eigenvalue method~\cite{Luscher:1990ck,Blossier:2009kd} can be
transferred and generalised to the other methods. In particular,
pollutions due to unresolved states decay exponentially in time.

At the beginning of the previous section we have demonstrated with
synthetic data that the PGEVM works as expected. In
particular, we could confirm that pollutions due to unresolved
excited states vanish exponentially in $t$. This exponential
convergence to the wanted state is faster if $\tMeff$ \eq{eq:Eeff} with $\delta t$ fixed is
used, as expected from the perturbative description. Increasing the
footprint of the Hankel matrix by increasing the parameter $\Delta$
helps in reducing the amplitude of the polluting terms. 

Still using synthetic data, we have shown that backward propagating
states affect PGEVM effective energies at large times. But, PGEVM
makes it also possible to distinguish forward from backward
propagating states.

As a first example for data with noise we have looked at the
pion. There are three important conclusions to be drawn here: first,
the PGEVM can also resolve forward and backward propagating states in
the presence of noise. Second, $\tMeff$ computed for fixed
$\delta t$ is advantageous compared to $\Meff$ at fixed
$t_0$, because in this case strong effects from the backward
propagating pion can be avoided.
And finally, combining GEVM and PGEVM sequentially leads to a
reduction of excited state contributions.

The next two QCD examples are for the $\eta$ meson and the $\rho$
meson where one must rely on the variational method. Moreover, the
signal to noise ratio decays exponentially such that excited state
reduction is imperative.

For the case of the $\eta$ meson the combined GEVM/PGEVM leads
to significantly larger confidence in the extracted energy levels. For the
$I=1, \pi-\pi$-scattering a strong improvement is visible. The latter is likely due
to the large input correlator matrix to the GEVM. This leads to a
large gap relevant for the corrections due to excited states and,
therefore, to small excited states in the PGEVM principal correlator.

Interestingly, for the $\rho$-meson example studied here also the
signal to noise ratio in the PGEVM principal correlator at fixed
$\delta t$ is competitive if not favourable compared to the effective
mass of the GEVM principal correlator.

Last but not least let us emphasise that the novel method presented here is not 
always advantageous and many other methods have been developed for the analysis 
of multi-exponential signals, each with their own strengths and weaknesses. We 
are especially referring to the recent developments of techniques based on the 
use of ordinary differential equations~\cite{ODE_methods} and the Gardner 
method~\cite{gardner_original}, for the latter see
appendix~\ref{sec:gardner}. Both methods are in principle capable of
extracting the full energy spectrum. However, the Gardner method 
becomes unreliable in the case of insufficient data and precision,
while we have not tested the ODE method here. But the results in
Ref.~\cite{ODE_methods} look promising.

\section{Summary}

In this paper we have clarified the relation among different methods
for the extraction of energy levels in lattice QCD available in the
literature. We have proposed and tested a new combination of
generalised eigenvalue and Prony method (GEVM/PGEVM), which helps to reduce excited state contaminations.

We have first discussed the systematic effects in the PGEVM stemming
from states not resolved by the method. They decay exponentially fast
in time with $\exp(-\Delta E_{n, l} t_0)$ with $\Delta E_{n, l}
=E_{n} -E_l$ the difference of the first not resolved energy level
$E_{n}$ and the level of interest $E_l$. Using synthetic data we
have shown that this is indeed the leading correction.

Next we have applied the method to a pion system and discussed its
ability to also determine backward propagating states, given high
enough statistical accuracy, see also
Ref.~\cite{Schiel:2015kwa}. Together with the results from the 
synthetic data we could also conclude that working at fixed
$\delta t$ is clearly advantageous compared to working at fixed $t_0$,
at least for data with little noise.

Finally, looking at lattice QCD examples for the $\eta$-meson and the
$\rho$-meson, we find that excited state contaminations can be reduced
significantly by using the combined GEVM/PGEVM. While it is not
clear whether also the statistical precision can be improved,
GEVM/PGEVM can significantly improve the confidence in the extraction
of energy levels, because plateaus start early enough in Euclidean
time. This is very much in line with the findings for the Prony method
in the version applied by the NPLQCD
collaboration~\cite{Beane:2009kya}.

The GEVM/PGEVM works particularly well, if in the first step
the GEVM removes as many intermediate states as possible and,
thus, the gap $\Delta E_{n, l}$ becomes as large as possible in the
PGEVM with moderately small $n$. The latter is important to
avoid numerical instabilities in the PGEVM.

~\\%
\begin{acknowledgments}
  The authors gratefully acknowledge the Gauss Centre for Supercomputing
  e.V. (www.gauss-centre.eu) for funding this project by providing
  computing time on the GCS Supercomputer JUQUEEN~\cite{juqueen} and the
  John von Neumann Institute for Computing (NIC) for computing time
  provided on the supercomputers JURECA~\cite{jureca} and JUWELS~\cite{juwels} at Jülich
  Supercomputing Centre (JSC).
  This project was funded in part by the DFG as a project in the
  Sino-German CRC110.
  The open source software packages tmLQCD~\cite{Jansen:2009xp,Abdel-Rehim:2013wba,Deuzeman:2013xaa}, 
  Lemon~\cite{Deuzeman:2011wz}, 
  QUDA~\cite{Clark:2009wm,Babich:2011np,Clark:2016rdz} and R~\cite{R:2019} have 
  been used.
\end{acknowledgments}

\appendix
\section{The Gardner method}
\label{sec:gardner}

	The Gardner method is a tool for the analysis of multicomponent exponential 
	decays. It completely avoids fits and uses Fourier transformations instead. 
	This global approach makes it extremely powerful, but also unstable. In 
	this section we discuss why we do not find the \gm\ applicable to 
	correlator analysis of lattice theories.

	\subsection{The algorithm}
	The most general form of a multicomponent exponential decaying function 
	$f(t)$ is
	\begin{align}
	f(t) &= \int_0^\infty g(\lambda)\eto{-\lambda 
		t}\,\md\lambda\label{eq:general_mult_exp_fun}
	\end{align}
	with some integrable function $g(\lambda)$ and $t$ bound from below, WLOG 
	$t\ge 0$. In the common discrete case we get
	\begin{align}
	g(\lambda) &= \sum_{i=0}^{\infty}A_i\delta(\lambda-E_i)
	\end{align}
	where the $A_i\in \mathbb R$ are the amplitudes, the $E_i$ are the decay 
	constants, often identified with energy levels, and $\delta$ denotes the 
	Dirac-Delta distribution. Gardner et al.~\cite{gardner_original} proposed 
	to multiply equation~\eqref{eq:general_mult_exp_fun} by $t=\exp(x)$ and 
	substitute $\lambda=\exp(-y)$ in order to obtain the convolution
	\begin{align}
	\eto{x}f\left(\eto{x}\right) &= 
		\int_{-\infty}^{\infty}g\left(\eto{-y}\right)\exp\left(-\eto{x-y}\right)\eto{x-y}\,\md 
		y\,.
	\end{align}
	This equation can now easily be solved for $g(\lambda)$ using Fourier 
	transformations. We define
	\begin{align}
	F(\mu) &\coloneqq 
		\frac{1}{\sqrt{2\pi}}\int_{-\infty}^{\infty}\eto{x}f\left(\eto{x}\right)\eto{\im 
		\mu x}\,\md x\,,\label{eq:ft_data_times_t}\\
	K(\mu) &\coloneqq 
		\frac{1}{\sqrt{2\pi}}\int_{-\infty}^{\infty}\exp\left(-\eto{x}\right)\eto{x}\eto{\im 
		\mu x}\,\md x\label{eq:ft_exp_e_e}\\
	&= \frac{1}{\sqrt{2\pi}}\Gamma(1+\im \mu)
	\end{align}
	and obtain
	\begin{align}
	g(\eto{-y}) &= 
		\frac{1}{2\pi}\int_{-\infty}^\infty\frac{F(\mu)}{K(\mu)}\eto{-\im 
		y\mu}\,\md \mu\,.\label{eq:ft_back_f_by_k}
	\end{align}
	The Fourier transformation in equation~\eqref{eq:ft_exp_e_e} has been 
	solved analytically, yielding the complex Gamma function $\Gamma$.

	The peaks of $g(\eto{-y})$ indicate the values of the $E_i$ by their 
	positions and the normalised amplitudes $A_i E_i$ by their heights. The 
	normalisation is due to the substitution 
	$g(\lambda)\mapsto\eto{-y}g(\eto{-y})$.

	\subsection{Numerical Precision}
	The Fourier integrals~\eqref{eq:ft_data_times_t} 
	and~\eqref{eq:ft_back_f_by_k} have to be solved numerically. We used the 
	extremely efficient algorithms \textit{double exponential 
	formulas}~\cite{dbl_exp_trafo} for low frequencies $\le 2\pi$ and 
	\textit{double exponential transformation for Fourier-type 
	integrals}~\cite{dbl_exp_osci} for high frequencies $\ge 2\pi$.

	These techniques allow to achieve machine precision of floating point 
	double precision arithmetics with $\lesssim 100$ function evaluations. This 
	however can only work as long as the result of the integral has the same 
	order of magnitude as the maximum of the integrated function. It turns out that 
	this is not the case for the given integrals. $F(\mu)$ decays exponentially 
	in $\ordnung{\exp(-\frac{\pi}{2}|\mu|)}$ (at the same rate as $K(\mu)$) if 
	$f(t)$ follows equation~\eqref{eq:general_mult_exp_fun}. Thus, as $|\mu|$ 
	grows, the sum of values $\eto{x}f\left(\eto{x}\right)\in\ordnung{1}$ 
	approaches zero more and more, loosing significant digits. To avoid this 
	effect one would have to employ higher precision arithmetics.

	With double precision arithmetics the values of $F(\mu)$ become completely 
	unreliable in the region $|\mu|\gtrsim 20$ where $F(\mu)$ approaches 
	machine precision. In practice we find that only $F(|\mu| \lesssim 10)$ is 
	precise enough to be trusted.
	\begin{figure}[htb]
		\centering
		\includegraphics[width=0.4\textwidth, page=2]{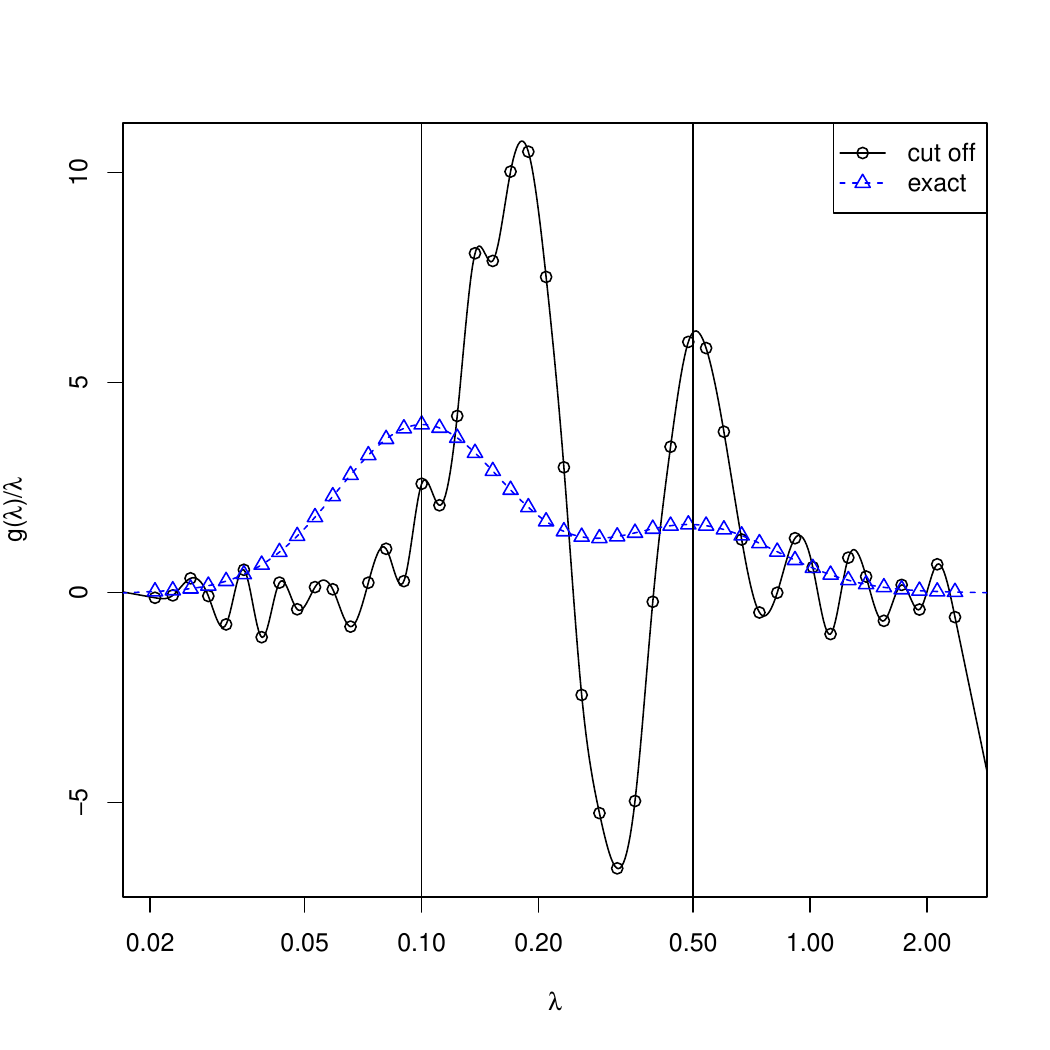}
		\caption{\gm\ applied to $f(t)=\eto{-\num{0.1}t}+2\eto{-\num{0.5}t}$. 
		Lines are cubic splines between the calculated points and guides to the 
		eye only. The black circles are obtained via a cubic spline to the 
		discrete set $\left\{\log(f(t))\,|\;t\in\{0,\dots,20\}\right\}$ with 
		$f(t>20)=0$. The blue triangles are obtained using the exact functional 
		form.}
		\label{fig:gardner_trafo_cut}
	\end{figure}

	\subsection{Limited data}
	In the case relevant for this work the data is limited to a noisy time 
	series $f(t)+\nu(t)$, $t\in\{0,\dots,n\}$, where $\nu(t)$ is an error. Thus 
	we have to deal with three difficulties, namely a discrete set, a finite 
	range and noise. Additional problems are the aforementioned limitation in 
	precision for high frequencies and possible small gaps between decay 
	constants $E_i$ that cannot be resolved. Ref.~\cite{gardner_review} 
	summarises a large number of improvements to the \gm\ and we are going to 
	mention the relevant ones explicitly below.

	\paragraph{Limited precision of $F(\mu)$}
	at high frequencies leads to a divergence of $\frac{F(\mu)}{K(\mu)}$ and 
	thus to a divergent integral in equation~\eqref{eq:ft_back_f_by_k}. If one 
	does not have or want to spend the resources for arbitrary precision 
	arithmetics, one is therefore forced to dampen the integrand 
	in~\eqref{eq:ft_back_f_by_k}. Gardner et al.~\cite{gardner_original} 
	originally proposed to simply introduce a cut off to the integral. It turns 
	out that this cut off leads to sinc-like oscillations of $g(\eto{-y})$, 
	i.e.\@ a high number of slowly decaying spurious peaks. These oscillations 
	can be removed by introducing a convergence factor of the form 
	$\exp(-\frac{\mu^2}{2w^2})$ instead of the cut off~\cite{gardner_gauss}. 
	The effective convolution of the exact result $g(\eto{-y})$ with a Gaussian 
	only smoothes $g(\eto{-y})$ but does not introduce oscillations. We chose 
	$w=2$ for our test runs. This choice does not always yield optimal results, 
	but it is very stable.

        \begin{figure}[htb]
	  \centering
	  \includegraphics[width=0.4\textwidth, page=4]{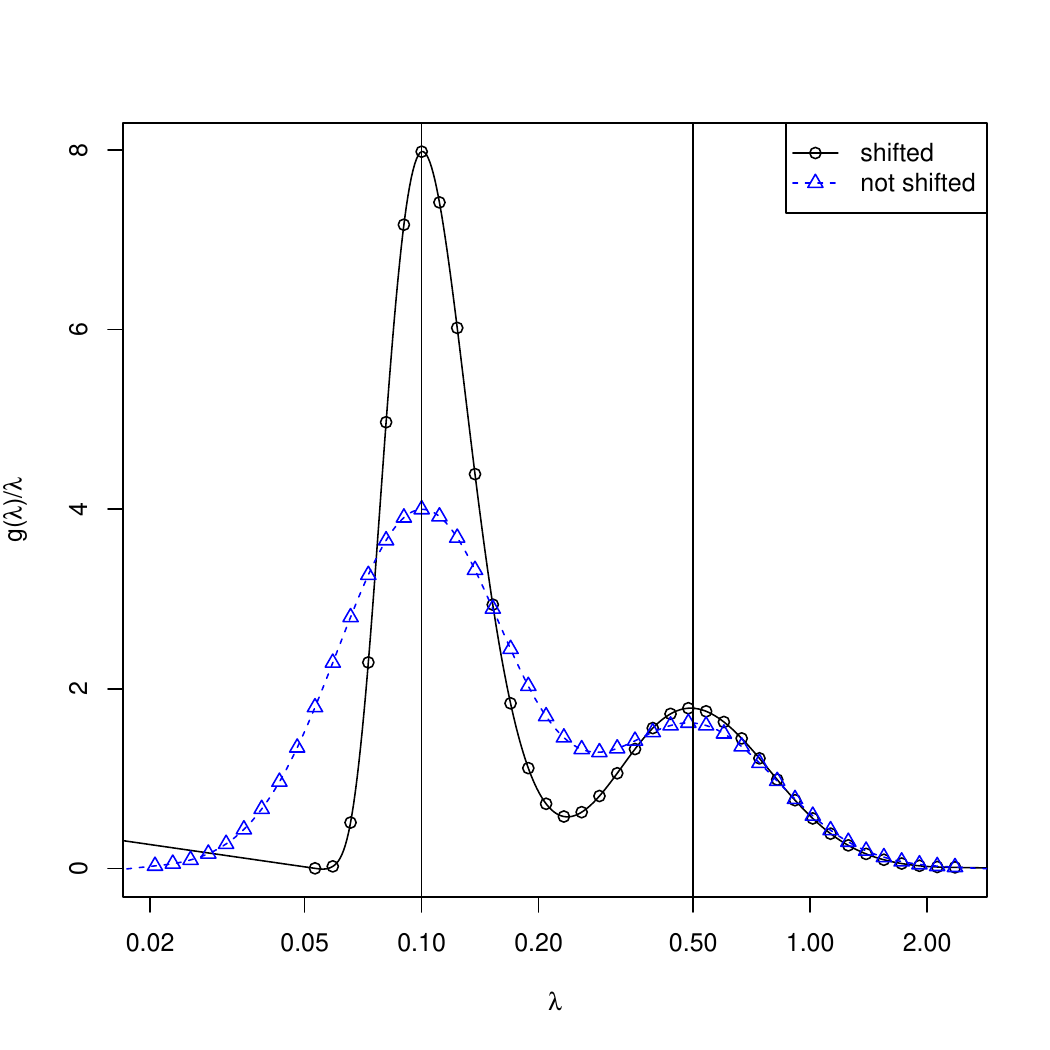}
	  \caption{\gm\ applied to $f(t)=\eto{-\num{0.1}t}+2\eto{-\num{0.5}t}$. 
	    Lines are cubic splines between the calculated points and guides to the 
	    eye only. The black circles are obtained via a cubic spline to the 
	    discrete set 
	    $\left\{\log(f(t)\eto{\num{.05}t})\,|\;t\in\{0,\dots,20\}\right\}$ with 
	    linear extrapolation. The blue triangles are obtained using the exact 
	    functional form.}
	  \label{fig:gardner_trafo_shift}
          
	\end{figure}

	\paragraph{Discrete data}
	is probably easiest to compensate. The exponential of a cubic spline of 
	$\log(f(t))$ yields a very precise interpolation of the data. Typically for 
	test functions the relative error is less than $10^{-4}$. Usually this is 
	far below noise level.

        \begin{figure}[htb]
	  \centering
	  \includegraphics[width=0.4\textwidth, page=3]{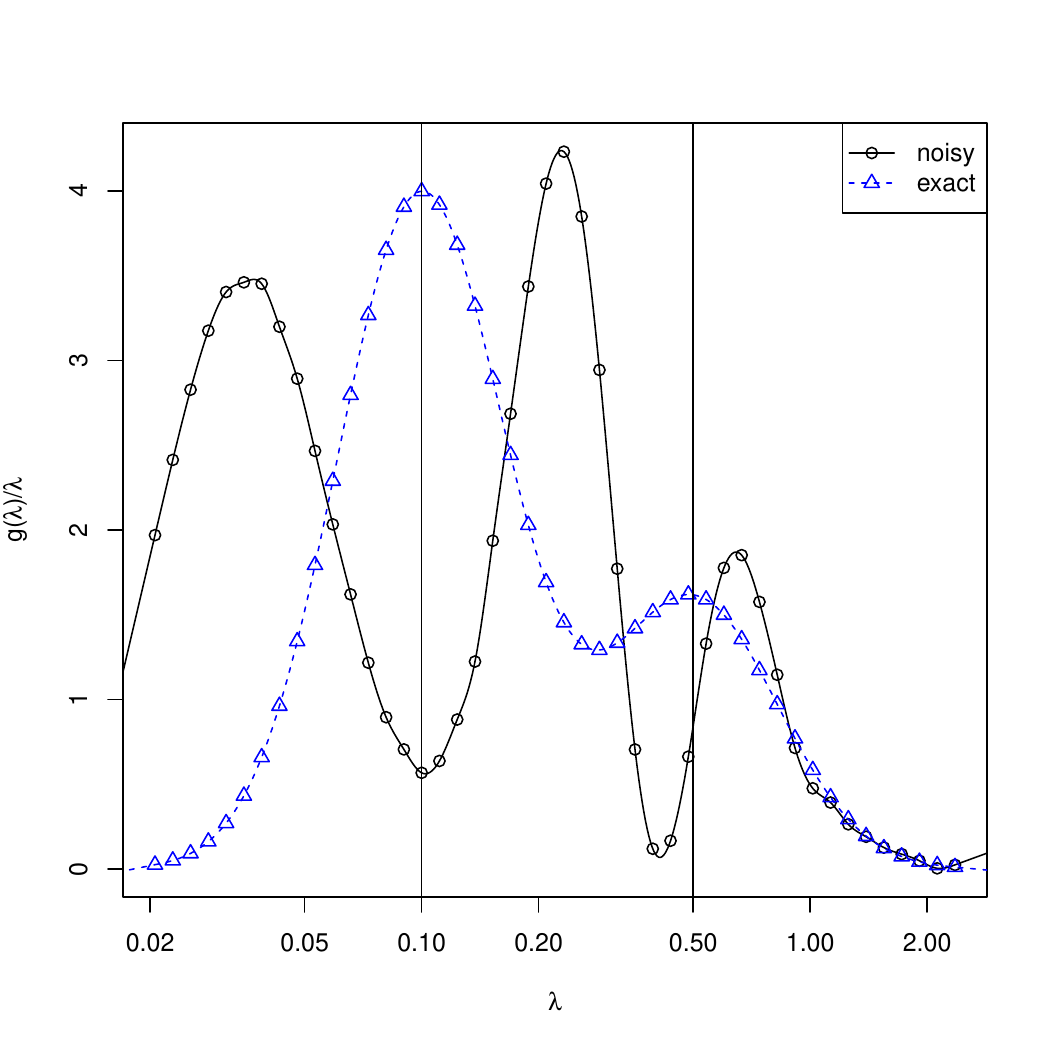}
	  \caption{\gm\ applied to $f(t)=\eto{-\num{0.1}t}+2\eto{-\num{0.5}t}$. 
	    Lines are cubic splines between the calculated points and guides to the 
	    eye only. The black circles are obtained via a cubic spline to the 
	    discrete set 
	    $\left\{\log(f(t))\,|\;t\in\{0,\dots,19\}\right\}\cup\left\{\log\left(\frac{f(19)+f(20)}{2}\right)\right\}$ 
	    with linear extrapolation. The blue triangles are obtained using the 
	    exact functional form.}
	  \label{fig:gardner_trafo_noise}
	\end{figure}

	\paragraph{Finite time range}
	is a much more severe problem. The exponential tail of $f(t)$ for 
	$t\rightarrow\infty$ carries a lot of information, especially about the 
	lowest decay modes. Thus extrapolation of the data essentially fixes the 
	ground state energy which we are usually most interested in. An 
	extrapolation of some kind is necessary, as a cut off completely obscures 
	the result (see \fig{fig:gardner_trafo_cut}). For a proper 
	extrapolation one would need to know at least the smallest $E_i$ in 
	advance, removing the necessity to apply the \gm\ in the first place. In 
	our test runs we used a linear extrapolation of the splines to the 
	log-data.

	Provencher~\cite{gardner_damped} proposes to multiply the complete time 
	series by a damping term of the form $t^\alpha\eto{-\beta t}$ with 
	$\alpha,\beta>0$ instead of $t$. This leads to a suppression of the region 
	beyond the data range, but it also moves the peaks of $g(\eto{-y})$ closer 
	together, thus decreasing the resolution. Still, Provencher does not remove 
	the necessity of an extrapolation completely. In addition the method 
	introduces two parameters that have to be tuned.

	Let us remark here that, given a reliable extrapolation or very long 
	measurement, the inverse of Provencher's method can be used to improve 
	resolution: Choose $\min(E_i) < \beta < 0$ and so separate the lowest lying 
	peak from the others. We show the advantage of such a shift of the decay 
	constants in \fig{fig:gardner_trafo_shift}.

	\paragraph{Noisy data}
	is not a significant problem by itself, as long as the magnitude is known. 
	Fluctuations can be captured by the bootstrap or other error propagating 
	methods. Severe problems arise if noise is combined with the aforementioned 
	finite range. Then extrapolations based on the last few points (e.g.\@ with 
	the spline method) become very unreliable. We show this effect in 
	\fig{fig:gardner_trafo_noise} where we slightly increased the value 
	of the very last data point.

	\subsection{Applicability in practice}
\begin{figure}[htb]
		\centering
		\includegraphics[width=0.4\textwidth, page=5]{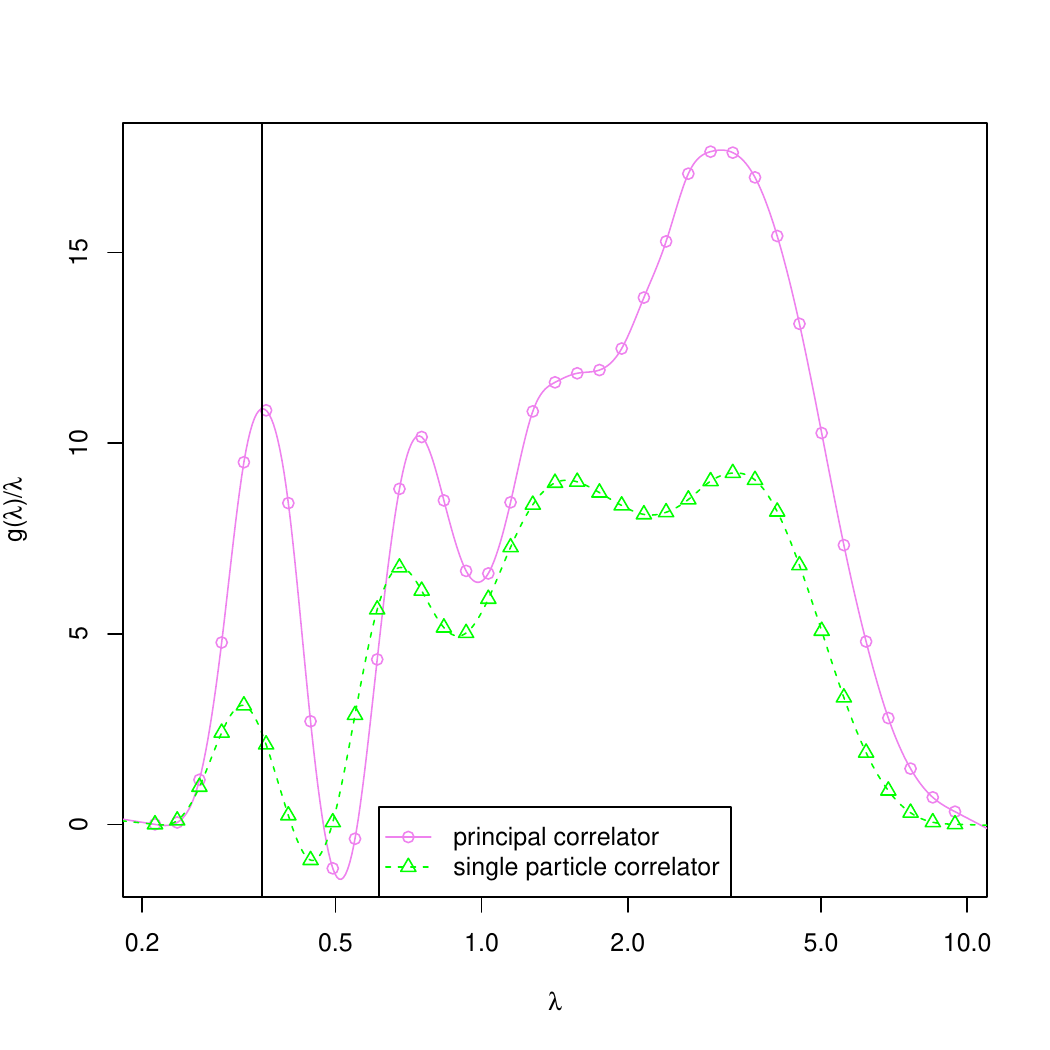}
		\caption{\gm\ with cubic spline inter- and extrapolation and 
		$\beta=\num{-.2}$ applied to the principal correlator obtained from a 
		gevp and the single particle correlator of a pion. Lines are cubic 
		splines between the calculated points and guides to the eye only. The 
		vertical line at $\lambda=\num{.353}$ shows the ground state obtained 
		from the principal correlator via to state $\chi^2$-fit.}
		\label{fig:gardner_trafo_data}
	\end{figure}

	We applied the method to some data obtained from lattice QCD simulations. 
	With some fine tuning of $\beta$ and a sensible truncation of the data (we 
	removed points below noise level and regions not falling monotonously) one 
	can obtain very good results. Note especially the high resolution of the 
	ground state in \fig{fig:gardner_trafo_data}, but the relevant 
	exited states can be resolved as well.

	Nevertheless we have to conclude that the \gm\ is not broadly applicable to 
	real data commonly obtained from lattice simulations. One reason is that it 
	requires fine tuning of several parameters to obtain good results. The main 
	problem however is the absence of a reliable extrapolation of noisy data 
	from the limited time range. The algorithm does not fail gracefully, i.e.\@ 
	there is no obvious check if the result for $g(\eto{-y})$ is correct or 
	not. Thus even though the \gm\ can yield very precise results, one cannot 
	automatise it and rely on the correctness of the output.

	As a last remark we would like to add that the \gm\ is also orders of 
	magnitude costlier in terms of computing resources than simpler methods 
	like $\chi^2$-fits.


\begin{thebibliography}{45}
\makeatletter
\providecommand \@ifxundefined [1]{
 \@ifx{#1\undefined}
}
\providecommand \@ifnum [1]{
 \ifnum #1\expandafter \@firstoftwo
 \else \expandafter \@secondoftwo
 \fi
}
\providecommand \@ifx [1]{
 \ifx #1\expandafter \@firstoftwo
 \else \expandafter \@secondoftwo
 \fi
}
\providecommand \natexlab [1]{#1}
\providecommand \enquote  [1]{``#1''}
\providecommand \bibnamefont  [1]{#1}
\providecommand \bibfnamefont [1]{#1}
\providecommand \citenamefont [1]{#1}
\providecommand \href@noop [0]{\@secondoftwo}
\providecommand \href [0]{\begingroup \@sanitize@url \@href}
\providecommand \@href[1]{\@@startlink{#1}\@@href}
\providecommand \@@href[1]{\endgroup#1\@@endlink}
\providecommand \@sanitize@url [0]{\catcode `\\12\catcode `\$12\catcode
  `\&12\catcode `\#12\catcode `\^12\catcode `\_12\catcode `\%12\relax}
\providecommand \@@startlink[1]{}
\providecommand \@@endlink[0]{}
\providecommand \url  [0]{\begingroup\@sanitize@url \@url }
\providecommand \@url [1]{\endgroup\@href {#1}{\urlprefix }}
\providecommand \urlprefix  [0]{URL }
\providecommand \Eprint [0]{\href }
\providecommand \doibase [0]{http://dx.doi.org/}
\providecommand \selectlanguage [0]{\@gobble}
\providecommand \bibinfo  [0]{\@secondoftwo}
\providecommand \bibfield  [0]{\@secondoftwo}
\providecommand \translation [1]{[#1]}
\providecommand \BibitemOpen [0]{}
\providecommand \bibitemStop [0]{}
\providecommand \bibitemNoStop [0]{.\EOS\space}
\providecommand \EOS [0]{\spacefactor3000\relax}
\providecommand \BibitemShut  [1]{\csname bibitem#1\endcsname}
\let\auto@bib@innerbib\@empty
\bibitem [{\citenamefont {Lepage}(1989)}]{Lepage:1989}
  \BibitemOpen
  \bibfield  {author} {\bibinfo {author} {\bibfnamefont {G.~P.}\ \bibnamefont
  {Lepage}},\ }\bibfield  {title} {\enquote {\bibinfo {title} {The analysis of
  algorithms for lattice field theory},}\ \ }(\bibinfo {year} {1989})\ \bibinfo
  {note} {invited lectures given at TASI’89 Summer School, Boulder, CO, Jun
  4-30, 1989. Published in Boulder ASI 1989:97-120
  (QCD161:T45:1989)}\BibitemShut {NoStop}
\bibitem [{\citenamefont {Feng}\ \emph {et~al.}(2011)\citenamefont {Feng},
  \citenamefont {Jansen},\ and\ \citenamefont {Renner}}]{Feng:2010es}
  \BibitemOpen
  \bibfield  {author} {\bibinfo {author} {\bibfnamefont {Xu}~\bibnamefont
  {Feng}}, \bibinfo {author} {\bibfnamefont {Karl}\ \bibnamefont {Jansen}}, \
  and\ \bibinfo {author} {\bibfnamefont {Dru~B.}\ \bibnamefont {Renner}},\
  }\bibfield  {title} {\enquote {\bibinfo {title} {{Resonance Parameters of the
  rho-Meson from Lattice QCD}},}\ }\href {\doibase 10.1103/PhysRevD.83.094505}
  {\bibfield  {journal} {\bibinfo  {journal} {Phys. Rev.}\ }\textbf {\bibinfo
  {volume} {D83}},\ \bibinfo {pages} {094505} (\bibinfo {year} {2011})},\
  \Eprint {http://arxiv.org/abs/1011.5288} {arXiv:1011.5288 [hep-lat]}
  \BibitemShut {NoStop}
\bibitem [{\citenamefont {Michael}\ and\ \citenamefont
  {Teasdale}(1983)}]{Michael:1982gb}
  \BibitemOpen
  \bibfield  {author} {\bibinfo {author} {\bibfnamefont {Christopher}\
  \bibnamefont {Michael}}\ and\ \bibinfo {author} {\bibfnamefont
  {I.}~\bibnamefont {Teasdale}},\ }\bibfield  {title} {\enquote {\bibinfo
  {title} {{Extracting Glueball Masses From Lattice {QCD}}},}\ }\href {\doibase
  10.1016/0550-3213(83)90674-0} {\bibfield  {journal} {\bibinfo  {journal}
  {Nucl. Phys.}\ }\textbf {\bibinfo {volume} {B215}},\ \bibinfo {pages}
  {433--446} (\bibinfo {year} {1983})}\BibitemShut {NoStop}
\bibitem [{\citenamefont {Lüscher}\ and\ \citenamefont
  {Wolff}(1990)}]{Luscher:1990ck}
  \BibitemOpen
  \bibfield  {author} {\bibinfo {author} {\bibfnamefont {Martin}\ \bibnamefont
  {Lüscher}}\ and\ \bibinfo {author} {\bibfnamefont {Ulli}\ \bibnamefont
  {Wolff}},\ }\bibfield  {title} {\enquote {\bibinfo {title} {{How to Calculate
  the Elastic Scattering Matrix in Two-dimensional Quantum Field Theories by
  Numerical Simulation}},}\ }\href {\doibase 10.1016/0550-3213(90)90540-T}
  {\bibfield  {journal} {\bibinfo  {journal} {Nucl. Phys.}\ }\textbf {\bibinfo
  {volume} {B339}},\ \bibinfo {pages} {222--252} (\bibinfo {year}
  {1990})}\BibitemShut {NoStop}
\bibitem [{\citenamefont {Blossier}\ \emph {et~al.}(2009)\citenamefont
  {Blossier}, \citenamefont {Della~Morte}, \citenamefont {von Hippel},
  \citenamefont {Mendes},\ and\ \citenamefont {Sommer}}]{Blossier:2009kd}
  \BibitemOpen
  \bibfield  {author} {\bibinfo {author} {\bibfnamefont {Benoit}\ \bibnamefont
  {Blossier}}, \bibinfo {author} {\bibfnamefont {Michele}\ \bibnamefont
  {Della~Morte}}, \bibinfo {author} {\bibfnamefont {Georg}\ \bibnamefont {von
  Hippel}}, \bibinfo {author} {\bibfnamefont {Tereza}\ \bibnamefont {Mendes}},
  \ and\ \bibinfo {author} {\bibfnamefont {Rainer}\ \bibnamefont {Sommer}},\
  }\bibfield  {title} {\enquote {\bibinfo {title} {{On the generalized
  eigenvalue method for energies and matrix elements in lattice field
  theory}},}\ }\href {\doibase 10.1088/1126-6708/2009/04/094} {\bibfield
  {journal} {\bibinfo  {journal} {JHEP}\ }\textbf {\bibinfo {volume} {04}},\
  \bibinfo {pages} {094} (\bibinfo {year} {2009})},\ \Eprint
  {http://arxiv.org/abs/0902.1265} {arXiv:0902.1265 [hep-lat]} \BibitemShut
  {NoStop}
\bibitem [{\citenamefont {de~Prony}(1795)}]{Prony:1795}
  \BibitemOpen
  \bibfield  {author} {\bibinfo {author} {\bibfnamefont {G.~R.}\ \bibnamefont
  {de~Prony}},\ }\href@noop {} {\bibfield  {journal} {\bibinfo  {journal}
  {Journal de l’cole Polytechnique}\ }\textbf {\bibinfo {volume} {1}},\
  \bibinfo {pages} {24--76} (\bibinfo {year} {1795})}\BibitemShut {NoStop}
\bibitem [{\citenamefont {Fleming}(2004)}]{Fleming:2004hs}
  \BibitemOpen
  \bibfield  {author} {\bibinfo {author} {\bibfnamefont {George~Tamminga}\
  \bibnamefont {Fleming}},\ }\bibfield  {title} {\enquote {\bibinfo {title}
  {{What can lattice QCD theorists learn from NMR spectroscopists?}}}\ }in\
  \href {http://www1.jlab.org/Ul/publications/view_pub.cfm?pub_id=5245} {\emph
  {\bibinfo {booktitle} {{QCD and numerical analysis III. Proceedings, 3rd
  International Workshop, Edinburgh, UK, June 30-July 4, 2003}}}}\ (\bibinfo
  {year} {2004})\ pp.\ \bibinfo {pages} {143--152},\ \Eprint
  {http://arxiv.org/abs/hep-lat/0403023} {arXiv:hep-lat/0403023 [hep-lat]}
  \BibitemShut {NoStop}
\bibitem [{\citenamefont {Beane}\ \emph {et~al.}(2009)\citenamefont {Beane},
  \citenamefont {Detmold}, \citenamefont {Luu}, \citenamefont {Orginos},
  \citenamefont {Parreno}, \citenamefont {Savage}, \citenamefont {Torok},\ and\
  \citenamefont {Walker-Loud}}]{Beane:2009kya}
  \BibitemOpen
  \bibfield  {author} {\bibinfo {author} {\bibfnamefont {Silas~R.}\
  \bibnamefont {Beane}}, \bibinfo {author} {\bibfnamefont {William}\
  \bibnamefont {Detmold}}, \bibinfo {author} {\bibfnamefont {Thomas~C.}\
  \bibnamefont {Luu}}, \bibinfo {author} {\bibfnamefont {Kostas}\ \bibnamefont
  {Orginos}}, \bibinfo {author} {\bibfnamefont {Assumpta}\ \bibnamefont
  {Parreno}}, \bibinfo {author} {\bibfnamefont {Martin~J.}\ \bibnamefont
  {Savage}}, \bibinfo {author} {\bibfnamefont {Aaron}\ \bibnamefont {Torok}}, \
  and\ \bibinfo {author} {\bibfnamefont {Andre}\ \bibnamefont {Walker-Loud}},\
  }\bibfield  {title} {\enquote {\bibinfo {title} {{High Statistics Analysis
  using Anisotropic Clover Lattices: (I) Single Hadron Correlation
  Functions}},}\ }\href {\doibase 10.1103/PhysRevD.79.114502} {\bibfield
  {journal} {\bibinfo  {journal} {Phys. Rev.}\ }\textbf {\bibinfo {volume}
  {D79}},\ \bibinfo {pages} {114502} (\bibinfo {year} {2009})},\ \Eprint
  {http://arxiv.org/abs/0903.2990} {arXiv:0903.2990 [hep-lat]} \BibitemShut
  {NoStop}
\bibitem [{\citenamefont {Fleming}\ \emph {et~al.}(2007)\citenamefont
  {Fleming}, \citenamefont {Cohen}, \citenamefont {Lin},\ and\ \citenamefont
  {Pereyra}}]{Fleming:2006zz}
  \BibitemOpen
  \bibfield  {author} {\bibinfo {author} {\bibfnamefont {George~T.}\
  \bibnamefont {Fleming}}, \bibinfo {author} {\bibfnamefont {Saul~D.}\
  \bibnamefont {Cohen}}, \bibinfo {author} {\bibfnamefont {Huey-Wen}\
  \bibnamefont {Lin}}, \ and\ \bibinfo {author} {\bibfnamefont {Victor}\
  \bibnamefont {Pereyra}},\ }\bibfield  {title} {\enquote {\bibinfo {title}
  {{Excited state effective masses}},}\ }\bibfield  {booktitle} {\emph
  {\bibinfo {booktitle} {{Proceedings, 25th International Symposium on Lattice
  field theory (Lattice 2007): Regensburg, Germany, July 30-August 4, 2007}}},\
  }\href {\doibase 10.22323/1.042.0096} {\bibfield  {journal} {\bibinfo
  {journal} {PoS}\ }\textbf {\bibinfo {volume} {LATTICE2007}},\ \bibinfo
  {pages} {096} (\bibinfo {year} {2007})}\BibitemShut {NoStop}
\bibitem [{\citenamefont {Berkowitz}\ \emph {et~al.}(2018)\citenamefont
  {Berkowitz}, \citenamefont {Nicholson}, \citenamefont {Chang}, \citenamefont
  {Rinaldi}, \citenamefont {Clark}, \citenamefont {Joó}, \citenamefont
  {Kurth}, \citenamefont {Vranas},\ and\ \citenamefont
  {Walker-Loud}}]{Berkowitz:2017smo}
  \BibitemOpen
  \bibfield  {author} {\bibinfo {author} {\bibfnamefont {Evan}\ \bibnamefont
  {Berkowitz}}, \bibinfo {author} {\bibfnamefont {Amy}\ \bibnamefont
  {Nicholson}}, \bibinfo {author} {\bibfnamefont {Chia~Cheng}\ \bibnamefont
  {Chang}}, \bibinfo {author} {\bibfnamefont {Enrico}\ \bibnamefont {Rinaldi}},
  \bibinfo {author} {\bibfnamefont {M.~A.}\ \bibnamefont {Clark}}, \bibinfo
  {author} {\bibfnamefont {Bálint}\ \bibnamefont {Joó}}, \bibinfo {author}
  {\bibfnamefont {Thorsten}\ \bibnamefont {Kurth}}, \bibinfo {author}
  {\bibfnamefont {Pavlos}\ \bibnamefont {Vranas}}, \ and\ \bibinfo {author}
  {\bibfnamefont {André}\ \bibnamefont {Walker-Loud}},\ }\bibfield  {title}
  {\enquote {\bibinfo {title} {{Calm Multi-Baryon Operators}},}\ }\bibfield
  {booktitle} {\emph {\bibinfo {booktitle} {{Proceedings, 35th International
  Symposium on Lattice Field Theory (Lattice 2017): Granada, Spain, June 18-24,
  2017}}},\ }\href {\doibase 10.1051/epjconf/201817505029} {\bibfield
  {journal} {\bibinfo  {journal} {EPJ Web Conf.}\ }\textbf {\bibinfo {volume}
  {175}},\ \bibinfo {pages} {05029} (\bibinfo {year} {2018})},\ \Eprint
  {http://arxiv.org/abs/1710.05642} {arXiv:1710.05642 [hep-lat]} \BibitemShut
  {NoStop}
\bibitem [{\citenamefont {Cushman}\ and\ \citenamefont
  {Fleming}(2019)}]{Cushman:2019hfh}
  \BibitemOpen
  \bibfield  {author} {\bibinfo {author} {\bibfnamefont {Kimmy~K.}\
  \bibnamefont {Cushman}}\ and\ \bibinfo {author} {\bibfnamefont {George~T.}\
  \bibnamefont {Fleming}},\ }\bibfield  {title} {\enquote {\bibinfo {title}
  {{Automated label flows for excited states of correlation functions in
  lattice gauge theory}},}\ }\href@noop {} {\  (\bibinfo {year} {2019})},\
  \Eprint {http://arxiv.org/abs/1912.08205} {arXiv:1912.08205 [hep-lat]}
  \BibitemShut {NoStop}
\bibitem [{\citenamefont {Banuls}\ \emph {et~al.}(2019)\citenamefont {Banuls},
  \citenamefont {Heller}, \citenamefont {Jansen}, \citenamefont {Knaute},\ and\
  \citenamefont {Svensson}}]{Banuls:2019qrq}
  \BibitemOpen
  \bibfield  {author} {\bibinfo {author} {\bibfnamefont {Mari~Carmen}\
  \bibnamefont {Banuls}}, \bibinfo {author} {\bibfnamefont {Michal~P.}\
  \bibnamefont {Heller}}, \bibinfo {author} {\bibfnamefont {Karl}\ \bibnamefont
  {Jansen}}, \bibinfo {author} {\bibfnamefont {Johannes}\ \bibnamefont
  {Knaute}}, \ and\ \bibinfo {author} {\bibfnamefont {Viktor}\ \bibnamefont
  {Svensson}},\ }\bibfield  {title} {\enquote {\bibinfo {title} {{From Spin
  Chains to Real-Time Thermal Field Theory Using Tensor Networks}},}\
  }\href@noop {} {\  (\bibinfo {year} {2019})},\ \Eprint
  {http://arxiv.org/abs/1912.08836} {arXiv:1912.08836 [hep-th]} \BibitemShut
  {NoStop}
\bibitem [{\citenamefont {Sauer}(2013)}]{sauer:2013}
  \BibitemOpen
  \bibfield  {author} {\bibinfo {author} {\bibfnamefont {Benedikt~Christian}\
  \bibnamefont {Sauer}},\ }\emph {\bibinfo {title} {Approaches to Improving
  $\eta^\prime$ Mass Calculations}},\ \href@noop {} {Master's thesis},\
  \bibinfo  {school} {University of Bonn} (\bibinfo {year} {2013})\BibitemShut
  {NoStop}
\bibitem [{\citenamefont {Irges}\ and\ \citenamefont
  {Knechtli}(2007)}]{Irges:2006hg}
  \BibitemOpen
  \bibfield  {author} {\bibinfo {author} {\bibfnamefont {Nikos}\ \bibnamefont
  {Irges}}\ and\ \bibinfo {author} {\bibfnamefont {Francesco}\ \bibnamefont
  {Knechtli}},\ }\bibfield  {title} {\enquote {\bibinfo {title} {{Lattice gauge
  theory approach to spontaneous symmetry breaking from an extra dimension}},}\
  }\href {\doibase 10.1016/j.nuclphysb.2007.01.023} {\bibfield  {journal}
  {\bibinfo  {journal} {Nucl. Phys.}\ }\textbf {\bibinfo {volume} {B775}},\
  \bibinfo {pages} {283--311} (\bibinfo {year} {2007})},\ \Eprint
  {http://arxiv.org/abs/hep-lat/0609045} {arXiv:hep-lat/0609045 [hep-lat]}
  \BibitemShut {NoStop}
\bibitem [{\citenamefont {Aubin}\ and\ \citenamefont
  {Orginos}(2011{\natexlab{a}})}]{Aubin:2010jc}
  \BibitemOpen
  \bibfield  {author} {\bibinfo {author} {\bibfnamefont {C.}~\bibnamefont
  {Aubin}}\ and\ \bibinfo {author} {\bibfnamefont {K.}~\bibnamefont
  {Orginos}},\ }\bibfield  {title} {\enquote {\bibinfo {title} {{A new approach
  for Delta form factors}},}\ }\bibfield  {booktitle} {\emph {\bibinfo
  {booktitle} {{Proceedings, 12th International Conference on Meson-nucleon
  physics and the structure of the nucleon (MENU 2000): Williamsburg, USA, May
  31-June 4, 2010}}},\ }\href {\doibase 10.1063/1.3647217} {\bibfield
  {journal} {\bibinfo  {journal} {AIP Conf. Proc.}\ }\textbf {\bibinfo {volume}
  {1374}},\ \bibinfo {pages} {621--624} (\bibinfo {year}
  {2011}{\natexlab{a}})},\ \Eprint {http://arxiv.org/abs/1010.0202}
  {arXiv:1010.0202 [hep-lat]} \BibitemShut {NoStop}
\bibitem [{\citenamefont {Aubin}\ and\ \citenamefont
  {Orginos}(2011{\natexlab{b}})}]{Aubin:2011zz}
  \BibitemOpen
  \bibfield  {author} {\bibinfo {author} {\bibfnamefont {C.}~\bibnamefont
  {Aubin}}\ and\ \bibinfo {author} {\bibfnamefont {K.}~\bibnamefont
  {Orginos}},\ }\bibfield  {title} {\enquote {\bibinfo {title} {{An improved
  method for extracting matrix elements from lattice three-point functions}},}\
  }\bibfield  {booktitle} {\emph {\bibinfo {booktitle} {{Proceedings, 29th
  International Symposium on Lattice field theory (Lattice 2011): Squaw Valley,
  Lake Tahoe, USA, July 10-16, 2011}}},\ }\href {\doibase 10.22323/1.139.0148}
  {\bibfield  {journal} {\bibinfo  {journal} {PoS}\ }\textbf {\bibinfo {volume}
  {LATTICE2011}},\ \bibinfo {pages} {148} (\bibinfo {year}
  {2011}{\natexlab{b}})}\BibitemShut {NoStop}
\bibitem [{\citenamefont {Schiel}(2015)}]{Schiel:2015kwa}
  \BibitemOpen
  \bibfield  {author} {\bibinfo {author} {\bibfnamefont {Rainer~W.}\
  \bibnamefont {Schiel}},\ }\bibfield  {title} {\enquote {\bibinfo {title}
  {{Expanding the Interpolator Basis in the Variational Method to Explicitly
  Account for Backward Running States}},}\ }\href {\doibase
  10.1103/PhysRevD.92.034512} {\bibfield  {journal} {\bibinfo  {journal} {Phys.
  Rev.}\ }\textbf {\bibinfo {volume} {D92}},\ \bibinfo {pages} {034512}
  (\bibinfo {year} {2015})},\ \Eprint {http://arxiv.org/abs/1503.02588}
  {arXiv:1503.02588 [hep-lat]} \BibitemShut {NoStop}
\bibitem [{\citenamefont {Ottnad}\ \emph {et~al.}(2018)\citenamefont {Ottnad},
  \citenamefont {Harris}, \citenamefont {Meyer}, \citenamefont {von Hippel},
  \citenamefont {Wilhelm},\ and\ \citenamefont {Wittig}}]{Ottnad:2017mzd}
  \BibitemOpen
  \bibfield  {author} {\bibinfo {author} {\bibfnamefont {Konstantin}\
  \bibnamefont {Ottnad}}, \bibinfo {author} {\bibfnamefont {Tim}\ \bibnamefont
  {Harris}}, \bibinfo {author} {\bibfnamefont {Harvey}\ \bibnamefont {Meyer}},
  \bibinfo {author} {\bibfnamefont {Georg}\ \bibnamefont {von Hippel}},
  \bibinfo {author} {\bibfnamefont {Jonas}\ \bibnamefont {Wilhelm}}, \ and\
  \bibinfo {author} {\bibfnamefont {Hartmut}\ \bibnamefont {Wittig}},\
  }\bibfield  {title} {\enquote {\bibinfo {title} {{Nucleon average quark
  momentum fraction with $N_\mathrm{f}=2+1$ Wilson fermions}},}\ }\bibfield
  {booktitle} {\emph {\bibinfo {booktitle} {{Proceedings, 35th International
  Symposium on Lattice Field Theory (Lattice 2017): Granada, Spain, June 18-24,
  2017}}},\ }\href {\doibase 10.1051/epjconf/201817506026} {\bibfield
  {journal} {\bibinfo  {journal} {EPJ Web Conf.}\ }\textbf {\bibinfo {volume}
  {175}},\ \bibinfo {pages} {06026} (\bibinfo {year} {2018})},\ \Eprint
  {http://arxiv.org/abs/1710.07816} {arXiv:1710.07816 [hep-lat]} \BibitemShut
  {NoStop}
\bibitem [{\citenamefont {Kostrzewa}\ \emph {et~al.}(2020)\citenamefont
  {Kostrzewa}, \citenamefont {Ostmeyer}, \citenamefont {Ueding},\ and\
  \citenamefont {Urbach}}]{hadron:2020}
  \BibitemOpen
  \bibfield  {author} {\bibinfo {author} {\bibfnamefont {Bartosz}\ \bibnamefont
  {Kostrzewa}}, \bibinfo {author} {\bibfnamefont {Johann}\ \bibnamefont
  {Ostmeyer}}, \bibinfo {author} {\bibfnamefont {Martin}\ \bibnamefont
  {Ueding}}, \ and\ \bibinfo {author} {\bibfnamefont {Carsten}\ \bibnamefont
  {Urbach}},\ }\href {https://github.com/HISKP-LQCD/hadron} {\enquote {\bibinfo
  {title} {hadron: package to extract hadronic quantities},}\ }\bibinfo
  {howpublished} {https://github.com/HISKP-LQCD/hadron} (\bibinfo {year}
  {2020}),\ \bibinfo {note} {{R} package version 3.0.1}\BibitemShut {NoStop}
\bibitem [{\citenamefont {Baron}\ \emph {et~al.}(2010)\citenamefont {Baron}
  \emph {et~al.}}]{Baron:2010bv}
  \BibitemOpen
  \bibfield  {author} {\bibinfo {author} {\bibfnamefont {R.}~\bibnamefont
  {Baron}} \emph {et~al.} (\bibinfo {collaboration} {ETM}),\ }\bibfield
  {title} {\enquote {\bibinfo {title} {{Light hadrons from lattice QCD with
  light (u,d), strange and charm dynamical quarks}},}\ }\href {\doibase
  10.1007/JHEP06(2010)111} {\bibfield  {journal} {\bibinfo  {journal} {JHEP}\
  }\textbf {\bibinfo {volume} {06}},\ \bibinfo {pages} {111} (\bibinfo {year}
  {2010})},\ \Eprint {http://arxiv.org/abs/1004.5284} {arXiv:1004.5284
  [hep-lat]} \BibitemShut {NoStop}
\bibitem [{\citenamefont {Boucaud}\ \emph {et~al.}(2008)\citenamefont {Boucaud}
  \emph {et~al.}}]{Boucaud:2008xu}
  \BibitemOpen
  \bibfield  {author} {\bibinfo {author} {\bibfnamefont {Philippe}\
  \bibnamefont {Boucaud}} \emph {et~al.} (\bibinfo {collaboration} {ETM}),\
  }\bibfield  {title} {\enquote {\bibinfo {title} {{Dynamical Twisted Mass
  Fermions with Light Quarks: Simulation and Analysis Details}},}\ }\href
  {\doibase 10.1016/j.cpc.2008.06.013} {\bibfield  {journal} {\bibinfo
  {journal} {Comput. Phys. Commun.}\ }\textbf {\bibinfo {volume} {179}},\
  \bibinfo {pages} {695--715} (\bibinfo {year} {2008})},\ \Eprint
  {http://arxiv.org/abs/0803.0224} {arXiv:0803.0224 [hep-lat]} \BibitemShut
  {NoStop}
\bibitem [{\citenamefont {Ottnad}\ and\ \citenamefont
  {Urbach}(2018)}]{Ottnad:2017bjt}
  \BibitemOpen
  \bibfield  {author} {\bibinfo {author} {\bibfnamefont {Konstantin}\
  \bibnamefont {Ottnad}}\ and\ \bibinfo {author} {\bibfnamefont {Carsten}\
  \bibnamefont {Urbach}} (\bibinfo {collaboration} {ETM}),\ }\bibfield  {title}
  {\enquote {\bibinfo {title} {{Flavor-singlet meson decay constants from
  $N_f=2+1+1$ twisted mass lattice QCD}},}\ }\href {\doibase
  10.1103/PhysRevD.97.054508} {\bibfield  {journal} {\bibinfo  {journal} {Phys.
  Rev.}\ }\textbf {\bibinfo {volume} {D97}},\ \bibinfo {pages} {054508}
  (\bibinfo {year} {2018})},\ \Eprint {http://arxiv.org/abs/1710.07986}
  {arXiv:1710.07986 [hep-lat]} \BibitemShut {NoStop}
\bibitem [{\citenamefont {Ottnad}\ \emph {et~al.}(2012)\citenamefont {Ottnad},
  \citenamefont {Michael}, \citenamefont {Reker}, \citenamefont {Urbach},
  \citenamefont {Michael}, \citenamefont {Reker},\ and\ \citenamefont
  {Urbach}}]{Ottnad:2012fv}
  \BibitemOpen
  \bibfield  {author} {\bibinfo {author} {\bibfnamefont {Konstantin}\
  \bibnamefont {Ottnad}}, \bibinfo {author} {\bibfnamefont {C.}~\bibnamefont
  {Michael}}, \bibinfo {author} {\bibfnamefont {S.}~\bibnamefont {Reker}},
  \bibinfo {author} {\bibfnamefont {C.}~\bibnamefont {Urbach}}, \bibinfo
  {author} {\bibfnamefont {Chris}\ \bibnamefont {Michael}}, \bibinfo {author}
  {\bibfnamefont {Siebren}\ \bibnamefont {Reker}}, \ and\ \bibinfo {author}
  {\bibfnamefont {Carsten}\ \bibnamefont {Urbach}} (\bibinfo {collaboration}
  {ETM}),\ }\bibfield  {title} {\enquote {\bibinfo {title} {{$\eta$ and $\eta'$
  mesons from $N_f=2+1+1$ twisted mass lattice QCD}},}\ }\href {\doibase
  10.1007/JHEP11(2012)048} {\bibfield  {journal} {\bibinfo  {journal} {JHEP}\
  }\textbf {\bibinfo {volume} {11}},\ \bibinfo {pages} {048} (\bibinfo {year}
  {2012})},\ \Eprint {http://arxiv.org/abs/1206.6719} {arXiv:1206.6719
  [hep-lat]} \BibitemShut {NoStop}
\bibitem [{\citenamefont {Michael}\ \emph {et~al.}(2013)\citenamefont
  {Michael}, \citenamefont {Ottnad},\ and\ \citenamefont
  {Urbach}}]{Michael:2013gka}
  \BibitemOpen
  \bibfield  {author} {\bibinfo {author} {\bibfnamefont {Chris}\ \bibnamefont
  {Michael}}, \bibinfo {author} {\bibfnamefont {Konstantin}\ \bibnamefont
  {Ottnad}}, \ and\ \bibinfo {author} {\bibfnamefont {Carsten}\ \bibnamefont
  {Urbach}} (\bibinfo {collaboration} {ETM}),\ }\bibfield  {title} {\enquote
  {\bibinfo {title} {{$\eta$ and $\eta^\prime$ mixing from Lattice QCD}},}\
  }\href {\doibase 10.1103/PhysRevLett.111.181602} {\bibfield  {journal}
  {\bibinfo  {journal} {Phys. Rev. Lett.}\ }\textbf {\bibinfo {volume} {111}},\
  \bibinfo {pages} {181602} (\bibinfo {year} {2013})},\ \Eprint
  {http://arxiv.org/abs/1310.1207} {arXiv:1310.1207 [hep-lat]} \BibitemShut
  {NoStop}
\bibitem [{\citenamefont {Werner}\ \emph {et~al.}(2019)\citenamefont {Werner}
  \emph {et~al.}}]{Werner:2019hxc}
  \BibitemOpen
  \bibfield  {author} {\bibinfo {author} {\bibfnamefont {Markus}\ \bibnamefont
  {Werner}} \emph {et~al.},\ }\bibfield  {title} {\enquote {\bibinfo {title}
  {{Hadron-Hadron Interactions from $N_f=2+1+1$ Lattice QCD: The
  $\rho$-resonance}},}\ }\href@noop {} {\  (\bibinfo {year} {2019})},\ \Eprint
  {http://arxiv.org/abs/1907.01237} {arXiv:1907.01237 [hep-lat]} \BibitemShut
  {NoStop}
\bibitem [{\citenamefont {Abdel-Rehim}\ \emph {et~al.}(2017)\citenamefont
  {Abdel-Rehim} \emph {et~al.}}]{Abdel-Rehim:2015pwa}
  \BibitemOpen
  \bibfield  {author} {\bibinfo {author} {\bibfnamefont {A.}~\bibnamefont
  {Abdel-Rehim}} \emph {et~al.} (\bibinfo {collaboration} {ETM}),\ }\bibfield
  {title} {\enquote {\bibinfo {title} {{First physics results at the physical
  pion mass from $N_f=2$ Wilson twisted mass fermions at maximal twist}},}\
  }\href {\doibase 10.1103/PhysRevD.95.094515} {\bibfield  {journal} {\bibinfo
  {journal} {Phys. Rev.}\ }\textbf {\bibinfo {volume} {D95}},\ \bibinfo {pages}
  {094515} (\bibinfo {year} {2017})},\ \Eprint
  {http://arxiv.org/abs/1507.05068} {arXiv:1507.05068 [hep-lat]} \BibitemShut
  {NoStop}
\bibitem [{\citenamefont {Liu}\ \emph {et~al.}(2017)\citenamefont {Liu} \emph
  {et~al.}}]{Liu:2016cba}
  \BibitemOpen
  \bibfield  {author} {\bibinfo {author} {\bibfnamefont {L.}~\bibnamefont
  {Liu}} \emph {et~al.},\ }\bibfield  {title} {\enquote {\bibinfo {title}
  {{Isospin-0 $\pi\pi$ s-wave scattering length from twisted mass lattice
  QCD}},}\ }\href {\doibase 10.1103/PhysRevD.96.054516} {\bibfield  {journal}
  {\bibinfo  {journal} {Phys. Rev.}\ }\textbf {\bibinfo {volume} {D96}},\
  \bibinfo {pages} {054516} (\bibinfo {year} {2017})},\ \Eprint
  {http://arxiv.org/abs/1612.02061} {arXiv:1612.02061 [hep-lat]} \BibitemShut
  {NoStop}
\bibitem [{\citenamefont {Romiti}\ and\ \citenamefont
  {Simula}(2019)}]{ODE_methods}
  \BibitemOpen
  \bibfield  {author} {\bibinfo {author} {\bibfnamefont {S.}~\bibnamefont
  {Romiti}}\ and\ \bibinfo {author} {\bibfnamefont {S.}~\bibnamefont
  {Simula}},\ }\bibfield  {title} {\enquote {\bibinfo {title} {{Extraction of
  multiple exponential signals from lattice correlation functions}},}\ }\href
  {\doibase 10.1103/PhysRevD.100.054515} {\bibfield  {journal} {\bibinfo
  {journal} {Phys. Rev. D}\ }\textbf {\bibinfo {volume} {100}},\ \bibinfo
  {pages} {054515} (\bibinfo {year} {2019})}\BibitemShut {NoStop}
\bibitem [{\citenamefont {Gardner}\ \emph {et~al.}(1959)\citenamefont
  {Gardner}, \citenamefont {Gardner}, \citenamefont {Laush},\ and\
  \citenamefont {Meinke}}]{gardner_original}
  \BibitemOpen
  \bibfield  {author} {\bibinfo {author} {\bibfnamefont {Donald~G.}\
  \bibnamefont {Gardner}}, \bibinfo {author} {\bibfnamefont {Jeanne~C.}\
  \bibnamefont {Gardner}}, \bibinfo {author} {\bibfnamefont {George}\
  \bibnamefont {Laush}}, \ and\ \bibinfo {author} {\bibfnamefont {W.~Wayne}\
  \bibnamefont {Meinke}},\ }\bibfield  {title} {\enquote {\bibinfo {title}
  {{Method for the Analysis of Multicomponent Exponential Decay Curves}},}\
  }\href {\doibase 10.1063/1.1730560} {\bibfield  {journal} {\bibinfo
  {journal} {The Journal of Chemical Physics}\ }\textbf {\bibinfo {volume}
  {31}},\ \bibinfo {pages} {978--986} (\bibinfo {year} {1959})},\ \Eprint
  {http://arxiv.org/abs/https://doi.org/10.1063/1.1730560}
  {https://doi.org/10.1063/1.1730560} \BibitemShut {NoStop}
\bibitem [{\citenamefont {{J\"{u}lich Supercomputing Centre}}(2015)}]{juqueen}
  \BibitemOpen
  \bibfield  {author} {\bibinfo {author} {\bibnamefont {{J\"{u}lich
  Supercomputing Centre}}},\ }\bibfield  {title} {\enquote {\bibinfo {title}
  {{JUQUEEN: IBM Blue Gene/Q Supercomputer System at the J\"{u}lich
  Supercomputing Centre}},}\ }\href {\doibase 10.17815/jlsrf-1-18} {\bibfield
  {journal} {\bibinfo  {journal} {Journal of large-scale research facilities}\
  }\textbf {\bibinfo {volume} {1}} (\bibinfo {year} {2015}),\
  10.17815/jlsrf-1-18}\BibitemShut {NoStop}
\bibitem [{\citenamefont {{J\"{u}lich Supercomputing Centre}}(2018)}]{jureca}
  \BibitemOpen
  \bibfield  {author} {\bibinfo {author} {\bibnamefont {{J\"{u}lich
  Supercomputing Centre}}},\ }\bibfield  {title} {\enquote {\bibinfo {title}
  {{JURECA: Modular supercomputer at J\"{u}lich Supercomputing Centre}},}\
  }\href {\doibase 10.17815/jlsrf-4-121-1} {\bibfield  {journal} {\bibinfo
  {journal} {Journal of large-scale research facilities}\ }\textbf {\bibinfo
  {volume} {4}} (\bibinfo {year} {2018}),\ 10.17815/jlsrf-4-121-1}\BibitemShut
  {NoStop}
\bibitem [{\citenamefont {{J\"{u}lich Supercomputing Centre}}(2019)}]{juwels}
  \BibitemOpen
  \bibfield  {author} {\bibinfo {author} {\bibnamefont {{J\"{u}lich
  Supercomputing Centre}}},\ }\bibfield  {title} {\enquote {\bibinfo {title}
  {{JUWELS: Modular Tier-0/1 Supercomputer at the J\"{u}lich Supercomputing
  Centre}},}\ }\href {\doibase 10.17815/jlsrf-5-171} {\bibfield  {journal}
  {\bibinfo  {journal} {Journal of large-scale research facilities}\ }\textbf
  {\bibinfo {volume} {5}} (\bibinfo {year} {2019}),\
  10.17815/jlsrf-5-171}\BibitemShut {NoStop}
\bibitem [{\citenamefont {Jansen}\ and\ \citenamefont
  {Urbach}(2009)}]{Jansen:2009xp}
  \BibitemOpen
  \bibfield  {author} {\bibinfo {author} {\bibfnamefont {K.}~\bibnamefont
  {Jansen}}\ and\ \bibinfo {author} {\bibfnamefont {C.}~\bibnamefont
  {Urbach}},\ }\bibfield  {title} {\enquote {\bibinfo {title} {{tmLQCD: A
  Program suite to simulate Wilson Twisted mass Lattice QCD}},}\ }\href
  {\doibase 10.1016/j.cpc.2009.05.016} {\bibfield  {journal} {\bibinfo
  {journal} {Comput.Phys.Commun.}\ }\textbf {\bibinfo {volume} {180}},\
  \bibinfo {pages} {2717--2738} (\bibinfo {year} {2009})},\ \Eprint
  {http://arxiv.org/abs/0905.3331} {arXiv:0905.3331 [hep-lat]} \BibitemShut
  {NoStop}
\bibitem [{\citenamefont {Abdel-Rehim}\ \emph {et~al.}(2014)\citenamefont
  {Abdel-Rehim}, \citenamefont {Burger}, \citenamefont {Deuzeman},
  \citenamefont {Jansen}, \citenamefont {Kostrzewa}, \citenamefont {Scorzato},\
  and\ \citenamefont {Urbach}}]{Abdel-Rehim:2013wba}
  \BibitemOpen
  \bibfield  {author} {\bibinfo {author} {\bibfnamefont {Abdou}\ \bibnamefont
  {Abdel-Rehim}}, \bibinfo {author} {\bibfnamefont {Florian}\ \bibnamefont
  {Burger}}, \bibinfo {author} {\bibfnamefont {Alber}\ \bibnamefont
  {Deuzeman}}, \bibinfo {author} {\bibfnamefont {Karl}\ \bibnamefont {Jansen}},
  \bibinfo {author} {\bibfnamefont {Bartosz}\ \bibnamefont {Kostrzewa}},
  \bibinfo {author} {\bibfnamefont {Luigi}\ \bibnamefont {Scorzato}}, \ and\
  \bibinfo {author} {\bibfnamefont {Carsten}\ \bibnamefont {Urbach}},\
  }\bibfield  {title} {\enquote {\bibinfo {title} {{Recent developments in the
  tmLQCD software suite}},}\ }\href {\doibase 10.22323/1.187.0414} {\bibfield
  {journal} {\bibinfo  {journal} {PoS}\ }\textbf {\bibinfo {volume}
  {LATTICE2013}},\ \bibinfo {pages} {414} (\bibinfo {year} {2014})},\ \Eprint
  {http://arxiv.org/abs/1311.5495} {arXiv:1311.5495 [hep-lat]} \BibitemShut
  {NoStop}
\bibitem [{\citenamefont {Deuzeman}\ \emph {et~al.}(2013)\citenamefont
  {Deuzeman}, \citenamefont {Jansen}, \citenamefont {Kostrzewa},\ and\
  \citenamefont {Urbach}}]{Deuzeman:2013xaa}
  \BibitemOpen
  \bibfield  {author} {\bibinfo {author} {\bibfnamefont {A.}~\bibnamefont
  {Deuzeman}}, \bibinfo {author} {\bibfnamefont {K.}~\bibnamefont {Jansen}},
  \bibinfo {author} {\bibfnamefont {B.}~\bibnamefont {Kostrzewa}}, \ and\
  \bibinfo {author} {\bibfnamefont {C.}~\bibnamefont {Urbach}},\ }\bibfield
  {title} {\enquote {\bibinfo {title} {{Experiences with OpenMP in tmLQCD}},}\
  }\href@noop {} {\bibfield  {journal} {\bibinfo  {journal} {PoS}\ }\textbf
  {\bibinfo {volume} {LATTICE2013}},\ \bibinfo {pages} {416} (\bibinfo {year}
  {2013})},\ \Eprint {http://arxiv.org/abs/1311.4521} {arXiv:1311.4521
  [hep-lat]} \BibitemShut {NoStop}
\bibitem [{\citenamefont {Deuzeman}\ \emph {et~al.}(2012)\citenamefont
  {Deuzeman}, \citenamefont {Reker},\ and\ \citenamefont
  {Urbach}}]{Deuzeman:2011wz}
  \BibitemOpen
  \bibfield  {author} {\bibinfo {author} {\bibfnamefont {Albert}\ \bibnamefont
  {Deuzeman}}, \bibinfo {author} {\bibfnamefont {Siebren}\ \bibnamefont
  {Reker}}, \ and\ \bibinfo {author} {\bibfnamefont {Carsten}\ \bibnamefont
  {Urbach}} (\bibinfo {collaboration} {ETM}),\ }\bibfield  {title} {\enquote
  {\bibinfo {title} {{Lemon: an MPI parallel I/O library for data encapsulation
  using LIME}},}\ }\href {\doibase 10.1016/j.cpc.2012.01.016} {\bibfield
  {journal} {\bibinfo  {journal} {Comput. Phys. Commun.}\ }\textbf {\bibinfo
  {volume} {183}},\ \bibinfo {pages} {1321--1335} (\bibinfo {year} {2012})},\
  \Eprint {http://arxiv.org/abs/1106.4177} {arXiv:1106.4177 [hep-lat]}
  \BibitemShut {NoStop}
\bibitem [{\citenamefont {Clark}\ \emph {et~al.}(2010)\citenamefont {Clark},
  \citenamefont {Babich}, \citenamefont {Barros}, \citenamefont {Brower},\ and\
  \citenamefont {Rebbi}}]{Clark:2009wm}
  \BibitemOpen
  \bibfield  {author} {\bibinfo {author} {\bibfnamefont {M.~A.}\ \bibnamefont
  {Clark}}, \bibinfo {author} {\bibfnamefont {R.}~\bibnamefont {Babich}},
  \bibinfo {author} {\bibfnamefont {K.}~\bibnamefont {Barros}}, \bibinfo
  {author} {\bibfnamefont {R.~C.}\ \bibnamefont {Brower}}, \ and\ \bibinfo
  {author} {\bibfnamefont {C.}~\bibnamefont {Rebbi}},\ }\bibfield  {title}
  {\enquote {\bibinfo {title} {{Solving Lattice QCD systems of equations using
  mixed precision solvers on GPUs}},}\ }\href {\doibase
  10.1016/j.cpc.2010.05.002} {\bibfield  {journal} {\bibinfo  {journal}
  {Comput. Phys. Commun.}\ }\textbf {\bibinfo {volume} {181}},\ \bibinfo
  {pages} {1517--1528} (\bibinfo {year} {2010})},\ \Eprint
  {http://arxiv.org/abs/0911.3191} {arXiv:0911.3191 [hep-lat]} \BibitemShut
  {NoStop}
\bibitem [{\citenamefont {Babich}\ \emph {et~al.}(2011)\citenamefont {Babich},
  \citenamefont {Clark}, \citenamefont {Joo}, \citenamefont {Shi},
  \citenamefont {Brower},\ and\ \citenamefont {Gottlieb}}]{Babich:2011np}
  \BibitemOpen
  \bibfield  {author} {\bibinfo {author} {\bibfnamefont {R.}~\bibnamefont
  {Babich}}, \bibinfo {author} {\bibfnamefont {M.~A.}\ \bibnamefont {Clark}},
  \bibinfo {author} {\bibfnamefont {B.}~\bibnamefont {Joo}}, \bibinfo {author}
  {\bibfnamefont {G.}~\bibnamefont {Shi}}, \bibinfo {author} {\bibfnamefont
  {R.~C.}\ \bibnamefont {Brower}}, \ and\ \bibinfo {author} {\bibfnamefont
  {S.}~\bibnamefont {Gottlieb}},\ }\bibfield  {title} {\enquote {\bibinfo
  {title} {{Scaling Lattice QCD beyond 100 GPUs}},}\ }in\ \href {\doibase
  10.1145/2063384.2063478} {\emph {\bibinfo {booktitle} {{SC11 International
  Conference for High Performance Computing, Networking, Storage and Analysis
  Seattle, Washington, November 12-18, 2011}}}}\ (\bibinfo {year} {2011})\
  \Eprint {http://arxiv.org/abs/1109.2935} {arXiv:1109.2935 [hep-lat]}
  \BibitemShut {NoStop}
\bibitem [{\citenamefont {Clark}\ \emph {et~al.}(2016)\citenamefont {Clark},
  \citenamefont {Joó}, \citenamefont {Strelchenko}, \citenamefont {Cheng},
  \citenamefont {Gambhir},\ and\ \citenamefont {Brower}}]{Clark:2016rdz}
  \BibitemOpen
  \bibfield  {author} {\bibinfo {author} {\bibfnamefont {M.~A.}\ \bibnamefont
  {Clark}}, \bibinfo {author} {\bibfnamefont {Bálint}\ \bibnamefont {Joó}},
  \bibinfo {author} {\bibfnamefont {Alexei}\ \bibnamefont {Strelchenko}},
  \bibinfo {author} {\bibfnamefont {Michael}\ \bibnamefont {Cheng}}, \bibinfo
  {author} {\bibfnamefont {Arjun}\ \bibnamefont {Gambhir}}, \ and\ \bibinfo
  {author} {\bibfnamefont {Richard}\ \bibnamefont {Brower}},\ }\bibfield
  {title} {\enquote {\bibinfo {title} {{Accelerating Lattice QCD Multigrid on
  GPUs Using Fine-Grained Parallelization}},}\ }\href@noop {} {\  (\bibinfo
  {year} {2016})},\ \Eprint {http://arxiv.org/abs/1612.07873} {arXiv:1612.07873
  [hep-lat]} \BibitemShut {NoStop}
\bibitem [{\citenamefont {{R Core Team}}(2019)}]{R:2019}
  \BibitemOpen
  \bibfield  {author} {\bibinfo {author} {\bibnamefont {{R Core Team}}},\
  }\href {https://www.R-project.org/} {\emph {\bibinfo {title} {R: A Language
  and Environment for Statistical Computing}}},\ \bibinfo {organization} {R
  Foundation for Statistical Computing},\ \bibinfo {address} {Vienna, Austria}
  (\bibinfo {year} {2019})\BibitemShut {NoStop}
\bibitem [{\citenamefont {Takahasi}\ and\ \citenamefont
  {Mori}(1973)}]{dbl_exp_trafo}
  \BibitemOpen
  \bibfield  {author} {\bibinfo {author} {\bibfnamefont {Hidetosi}\
  \bibnamefont {Takahasi}}\ and\ \bibinfo {author} {\bibfnamefont {Masatake}\
  \bibnamefont {Mori}},\ }\bibfield  {title} {\enquote {\bibinfo {title}
  {{Double Exponential Formulas for Numerical Integration}},}\ }\href {\doibase
  10.2977/prims/1195192451} {\bibfield  {journal} {\bibinfo  {journal}
  {Publications of the Research Institute for Mathematical Sciences}\ }\textbf
  {\bibinfo {volume} {9}},\ \bibinfo {pages} {721--741} (\bibinfo {year}
  {1973})}\BibitemShut {NoStop}
\bibitem [{\citenamefont {Ooura}\ and\ \citenamefont
  {Mori}(1999)}]{dbl_exp_osci}
  \BibitemOpen
  \bibfield  {author} {\bibinfo {author} {\bibfnamefont {Takuya}\ \bibnamefont
  {Ooura}}\ and\ \bibinfo {author} {\bibfnamefont {Masatake}\ \bibnamefont
  {Mori}},\ }\bibfield  {title} {\enquote {\bibinfo {title} {{A robust double
  exponential formula for Fourier-type integrals}},}\ }\href {\doibase
  https://doi.org/10.1016/S0377-0427(99)00223-X} {\bibfield  {journal}
  {\bibinfo  {journal} {Journal of Computational and Applied Mathematics}\
  }\textbf {\bibinfo {volume} {112}},\ \bibinfo {pages} {229 -- 241} (\bibinfo
  {year} {1999})}\BibitemShut {NoStop}
\bibitem [{\citenamefont {Jibia}\ and\ \citenamefont
  {Salami}(2012)}]{gardner_review}
  \BibitemOpen
  \bibfield  {author} {\bibinfo {author} {\bibfnamefont {Abdussamad}\
  \bibnamefont {Jibia}}\ and\ \bibinfo {author} {\bibfnamefont {Momoh}\
  \bibnamefont {Salami}},\ }\bibfield  {title} {\enquote {\bibinfo {title} {{An
  Appraisal of Gardner Transform-Based Methods of Transient Multiexponential
  Signal Analysis}},}\ }\href {\doibase 10.7763/IJCTE.2012.V4.420} {\bibfield
  {journal} {\bibinfo  {journal} {International Journal of Computer Theory and
  Engineering}\ }\textbf {\bibinfo {volume} {4}},\ \bibinfo {pages} {16--25}
  (\bibinfo {year} {2012})}\BibitemShut {NoStop}
\bibitem [{\citenamefont {{Cohn-Sfetcu}}\ \emph {et~al.}(1975)\citenamefont
  {{Cohn-Sfetcu}}, \citenamefont {{Smith}}, \citenamefont {{Nichols}},\ and\
  \citenamefont {{Henry}}}]{gardner_gauss}
  \BibitemOpen
  \bibfield  {author} {\bibinfo {author} {\bibfnamefont {S.}~\bibnamefont
  {{Cohn-Sfetcu}}}, \bibinfo {author} {\bibfnamefont {M.~R.}\ \bibnamefont
  {{Smith}}}, \bibinfo {author} {\bibfnamefont {S.~T.}\ \bibnamefont
  {{Nichols}}}, \ and\ \bibinfo {author} {\bibfnamefont {D.~L.}\ \bibnamefont
  {{Henry}}},\ }\bibfield  {title} {\enquote {\bibinfo {title} {A digital
  technique for analyzing a class of multicomponent signals},}\ }\href
  {\doibase 10.1109/PROC.1975.9975} {\bibfield  {journal} {\bibinfo  {journal}
  {Proceedings of the IEEE}\ }\textbf {\bibinfo {volume} {63}},\ \bibinfo
  {pages} {1460--1467} (\bibinfo {year} {1975})}\BibitemShut {NoStop}
\bibitem [{\citenamefont {Provencher}(1976)}]{gardner_damped}
  \BibitemOpen
  \bibfield  {author} {\bibinfo {author} {\bibfnamefont {S.W.}\ \bibnamefont
  {Provencher}},\ }\bibfield  {title} {\enquote {\bibinfo {title} {{A Fourier
  method for the analysis of exponential decay curves}},}\ }\href {\doibase
  https://doi.org/10.1016/S0006-3495(76)85660-3} {\bibfield  {journal}
  {\bibinfo  {journal} {Biophysical Journal}\ }\textbf {\bibinfo {volume}
  {16}},\ \bibinfo {pages} {27 -- 41} (\bibinfo {year} {1976})}\BibitemShut
  {NoStop}
\end{thebibliography}
\end{document}